\documentclass[pdflatex,sn-apa]{sn-jnl}

\usepackage{graphicx}
\usepackage{amsmath,amssymb,amsfonts,mathtools}
\usepackage{amsthm}
\usepackage{bm}
\usepackage{booktabs}
\usepackage{array}
\usepackage{enumitem}
\usepackage{xcolor}
\usepackage{tikz}
\usepackage{longtable} 
\usepackage[normalem]{ulem}
\usepackage[utf8]{inputenc}
\usepackage[T1]{fontenc}

\usetikzlibrary{positioning,arrows.meta,shapes,calc,decorations.pathmorphing,decorations.pathreplacing}
\usepackage[title]{appendix}

\numberwithin{equation}{section}
\theoremstyle{thmstyleone}
\newtheorem{theorem}{Theorem}[section]
\newtheorem{lemma}[theorem]{Lemma}
\newtheorem{proposition}[theorem]{Proposition}
\newtheorem{corollary}[theorem]{Corollary}

\theoremstyle{thmstylethree}
\newtheorem{definition}[theorem]{Definition}
\newtheorem{assumption}[theorem]{Assumption}

\theoremstyle{thmstyletwo}
\newtheorem{remark}[theorem]{Remark}

\providecommand{\Z}{\mathbb{Z}}
\providecommand{\Q}{\mathbb{Q}}
\providecommand{\R}{\mathbb{R}}
\providecommand{\N}{\mathbb{N}}

\DeclareMathOperator{\Diff}{Diff}

\title[Sensory Dynamics in Recurrent Networks]{Embedding of Low‑Dimensional Sensory Dynamics in Recurrent Networks: Implications for the Geometry of Neural Representation}

\author*[1]{\fnm{Vikas N.} \sur{O'Reilly-Shah}}\email{voreill@uw.edu}
\author[2,3]{\fnm{Alessandro Maria} \sur{Selvitella}}\email{aselvite@pfw.edu}

\affil*[1]{\orgdiv{Department of Anesthesiology \& Pain Medicine},
\orgname{University of Washington School of Medicine},
\orgaddress{\street{1959 NE Pacific St (RR450)}, \city{Seattle}, \state{WA}, \postcode{98195}, \country{USA}}}

\affil[2]{\orgdiv{Department of Mathematical Sciences \& Laboratory of Data Science},
\orgname{Purdue University Fort Wayne},
\orgaddress{\city{Fort Wayne}, \state{IN}, \country{USA}}}

\affil[3]{\orgdiv{eScience Institute},
\orgname{University of Washington},
\orgaddress{\city{Seattle}, \state{WA}, \country{USA}}}

\abstract{ 
Neural population activity in sensory cortex is organized on low-dimensional manifolds, but it is unclear why such manifolds should arise and what determines their geometry. We address this \emph{sensory representation problem} by modeling cortical populations as recurrent circuits driven by low-dimensional, regular sensory dynamics (e.g.\ motion on a circle, head direction, multi-frequency tones on tori). By combining tools from generalized synchronization and delay-embedding theory, specialized to this quasiperiodic regime, we show that contracting recurrent networks generically develop smooth internal manifolds that embed the sensory dynamics. The dimensional requirement is modest and depends only on the intrinsic dimension $d$ of the effective sensory manifold, not on the complexity of the external world: a hidden dimension $N>2d$ generically suffices (e.g.\ $N\ge 3$ for a circle, $N\ge 5$ for a two-frequency torus; bounds compatible with Whitney and Takens' embedding theorems).

We then prove a prediction--separation result that links representational geometry directly to predictive performance, without assuming knowledge of contraction rates: if the circuit can predict future sensory inputs with small error, then states with different futures must be separated in neural state space, up to a resolution set by the prediction error. The resulting scale-limited embeddings naturally give rise to categorical boundaries, metameric equivalence of distinct stimuli, and discrimination thresholds. 

Numerical experiments with trained $\tanh$ recurrent networks driven by head-direction-like and multi-frequency signals recover ring- and torus-shaped hidden manifolds with the expected topology; state separation improves most rapidly near the $2d+1$ threshold. Training typically pushes the networks beyond the strict contraction regime where the theory guarantees faithful embedding, yet convergence consistent with generalized synchronization and manifold recovery persist, indicating that our conditions are sufficient but not necessary. 

Together, these results provide a mechanistic account of why low-dimensional sensory manifolds emerge in recurrent circuits and how prediction constrains their resolution, grounded in dynamical systems embedding theory and consistent with empirical findings on cortical population dynamics.
} 


\begin{document}

\maketitle

\section{Introduction}
\label{sec:introduction}

\subsection{The Sensory Representation Problem}
\label{subsec:sensory_problem}
How do neural circuits build internal representations of the sensory world? This question sits at the intersection of three converging research programs.

First, recordings from large neural populations show that task-relevant activity is confined to low-dimensional subspaces of the high-dimensional space of possible firing patterns. Motor cortical activity during reaching occupies manifolds of dimension 10--15 \citep{ChurchlandEtAl2012}. Head-direction cells form ring-shaped manifolds encoding orientation \citep{Kim2017, Chaudhuri2019}. Grid cell populations exhibit toroidal organization \citep{Gardner2022}. These structures are preserved across behaviors \citep{Gallego2018} and stable over years \citep{Gallego2020}. The emerging view, termed the \emph{neural manifold hypothesis}\footnote{ A glossary of key terms for readers from different disciplinary backgrounds is provided in Appendix~\ref{app:glossary}, covering dynamical systems (e.g., attractor, quasiperiodic dynamics, diffeomorphism), reservoir computing and generalized synchronization (e.g., echo state property, bunching condition, synchronization function), neural networks (e.g., truncated backpropagation, operator norm, spectral radius), neural coding and population dynamics (e.g., neural manifold hypothesis, persistent homology, computation through dynamics), and computational neuroscience (e.g., categorical perception, metameric collapse, predictive processing).}, holds that population-level geometry is the natural level of description for neural representation \citep{CunninghamYu2014, Gallego2017, VyasEtAl2020}. These manifolds carry distinct topological signatures: a ring ($S^1$) for head direction \citep{Chaudhuri2019}, a torus ($T^2$) for grid cells \citep{Gardner2022}, a low-dimensional subspace for motor cortex \citep{Gallego2017}, and a smooth high-dimensional code for visual cortex \citep{Stringer2019Geom}. Why these manifolds form, and what determines their geometry, is debated. Continuous-attractor-network models reproduce the dynamics of the ring and the torus once the recurrent connectivity is in place, but they assume a translation-invariant connectivity rather than deriving the manifold from the sensory drive \citep{Burak2009, Khona2022}.

Second, a separate line of work proposes that cortical circuits build internal models of sensory dynamics and continuously update those models to minimize the discrepancy between predicted and observed signals. This framework, known as \emph{predictive processing}, treats sensory representation as active model construction rather than passive feature detection \citep{RaoBallard1999, Friston2010, clark2013}. Its mathematical formulation rests on Bayesian analysis of stochastic differential equations rather than on geometric or topological tools. 

Third, work in dynamical systems theory shows that recurrent networks driven by external signals can develop internal states that faithfully mirror the dynamics of the driving process. Two recent studies make this concrete. \citet{UribarriMindlin2022} trained Long Short-Term Memory (LSTM) networks to predict chaotic time series and found that 94\% of successful models developed hidden-state trajectories preserving the topological organization of the driving system; models that failed to do so had worse prediction error. \citet{Ostrow2024} found a strong correlation ($r = 0.76$ for nonlinear decoding) between prediction performance and embedding quality across multiple architectures. Recent mathematical results from \citet{Hart2025} and \citet{Duan2023} have provided a rigorous basis upon which to understand why such embeddings arise.  

These three programs converge on a shared question that none has answered individually. The neural manifold literature documents low-dimensional representational geometry but may not offer a satisfactory thesis as to how it arises. Predictive processing provides a normative account of sensory inference but does not derive the geometry of its representations from first principles or bridge it to environmental dynamics. The dynamical systems literature demonstrates that, under appropriate conditions, driven recurrent networks can embed their inputs; existing results address chaotic attractors where the conditions are difficult to verify. These latter results have yet to be applied to a biological context.

This paper addresses the gap by the integration of mathematical techniques from essentially two different sources.  In \emph{reservoir computing}, a fixed recurrent network is driven by input and only a readout layer is trained; the network's recurrent dynamics implicitly encode the input's temporal structure \citep{Jaeger2001, HartHookDawes2020, Hart2025}. When these driven recurrent networks forget their initial state and settle into a stable correspondence between driving states and neural states, that correspondence is called \emph{generalized synchronization}. When a scalar time series is reconstructed into a higher-dimensional state space using time-delayed copies of the observed variable, the technique is called \emph{delay-coordinate embedding}. We combine these tools and specialize them to a regime where the sensory dynamics are regular and low-dimensional: invariant circles ($S^1$) and tori ($T^k$) arising in motion tracking, pitch perception, rhythm, and head-direction coding (described in greater detail in Section~\ref{subsec:sensory_regular}). In this regime, we obtain explicit, biologically interpretable conditions for faithful representation.

\subsection{Mathematical Approach}
\label{subsec:math_approach}

Sensory cortical circuits can be modeled as driven recurrent systems whose population state evolves in response to sensory input:
\begin{equation}
\label{eq:rnn_abstract}
h_{t+1} = F(h_t, u_t),
\end{equation}
where $h_t \in \R^N$ is the neural population state (a vector of $N$ activation values), $u_t$ is the sensory signal, and $F:\R^N\times\R\to\R^N$ is the state-update map. When the input derives from observing an external dynamical process, we write $u_t = \omega(\phi^t(x_0))$, where $\phi$ describes the external dynamics and $\omega$ is the observation function. This is a \emph{skew-product system} \citep{stark1999regularity}: the environment shapes the neural state, but the neural state does not influence the environment.

\begin{definition}[Recurrent network]
\label{def:rnn}
A \emph{recurrent network} is a tuple $\mathcal{R}=(\mathcal{H},\mathcal{X},\mathcal{W},h_0,\theta,F)$, where $\mathcal{H}=\R^N$ is the hidden-state space, $\mathcal{X}\subseteq\R$ is the input space, $\mathcal{W}$ is the parameter space, $h_0\in\mathcal{H}$ is the initial hidden state, $\theta=(W,W^{\mathrm{in}},b)\in\mathcal{W}$ are the parameters, and $F:\mathcal{H}\times\mathcal{X}\to\mathcal{H}$ is the state-update map fixed by $\theta$. Given an input sequence $(u_1,u_2,\dots)$ with $u_t\in\mathcal{X}$, the hidden state evolves by $h_{t+1}=F(h_t,u_t)$ as in~\eqref{eq:rnn_abstract}. The concrete instance studied here is the $\tanh$ network
\begin{equation}
\label{eq:rnn_tanh}
h_{t+1}=\tanh\!\big(W h_t+W^{\mathrm{in}}u_t+b\big),
\end{equation}
with recurrent weights $W$, input weights $W^{\mathrm{in}}$, and bias $b$. For the prediction task of Section~\ref{sec:dim_pred_geometry}, a readout $P:\mathcal{H}\to\R^{K+1}$ maps the hidden state to predicted future inputs.
\end{definition}
\par

Under suitable contraction conditions, such a driven network settles into a stable correspondence between driving states and neural states, described by a \emph{synchronization function} $f: M \to \R^N$, where $M$ is the compact manifold of environmental states formalized in Section~\ref{sec:preliminaries} \citep{HartHookDawes2020, Hart2025}. The network's hidden state encodes a compressed record of recent input history, playing the role that an explicit delay vector plays in classical delay-coordinate embedding \citep{Takens1981, SauerYorkeCasdagli1991}. Section~\ref{sec:preliminaries} defines these objects precisely.

These tools sit in an established lineage of forced-system embedding. Takens' theorem and its measure-theoretic refinement (Sauer, Yorke, and Casdagli) treat autonomous reconstruction. \citet{Stark1999delay} extended delay embedding to externally forced systems, and in a companion analysis showed that contracting forced systems possess a globally attracting invariant graph, the synchronization function, that is smooth under sufficiently strong contraction \citep{stark1999regularity}. Generalized synchronization and the reservoir-computing embedding results of \citet{Hart2025} and \citet{Duan2023} are the modern descendants. We follow this lineage: the sensory state is the forcing system, the recurrent population is the driven system, and the synchronization function is the invariant graph.

Several questions arise about this correspondence. If $f$ fails to be one-to-one, distinct environmental states collapse onto the same neural state, and the circuit cannot distinguish them. If $f$ is continuous but rough, small changes in the environment could produce arbitrarily oriented changes in the neural state, distorting the geometric relationships the circuit tracks. A natural hierarchy of representational quality therefore presents itself: continuity (nearby states map to nearby states), injectivity (distinct states remain distinct), and smoothness (local geometric structure is preserved). A $C^1$ embedding (a smooth, injective map with smooth inverse on its image) achieves all three. Whether a given circuit attains any of these levels depends on the interaction between the network's dynamics and the driving signal. The generalized synchronization framework provides one set of sufficient conditions, developed in Section~\ref{sec:main_results} (smoothness) and Section~\ref{sec:dimensions} (injectivity). These conditions are explicit and checkable in the regular regime, but they are not the only route to faithful representation; the numerical experiments in Section~\ref{subsec:numerical_experiments} confirm that embedding persists well beyond the regime where the sufficient conditions hold.

Within the generalized synchronization framework, a sufficient condition for smoothness is that the circuit damps perturbations faster than the driving dynamics expands them when traced backward. This is the \emph{bunching condition} \citep{stark1999regularity}; when it holds, the synchronization function is $C^1$, meaning it has a continuous first derivative and therefore preserves tangent-space structure. When the synchronization function is $C^1$, infinitesimal directions in the driving manifold map to infinitesimal directions in neural state space. This is the minimum needed for the induced dynamics on the image $f(M)$ to be a diffeomorphism, a smooth invertible map (Theorem~\ref{thm:main_synthesis}(iii)). Contraction is not the only route to diffeomorphism. As a matter of general differential topology, a diffeomorphism and its inverse cannot both be contractions \citep{Spivak1971-ux}. Contraction yields explicit inequalities in the driven recurrent setting. A separate result, developed in Section~\ref{sec:dim_pred_geometry}, constrains representational geometry through prediction accuracy without assuming contraction at all.

\subsection{Contributions}

This paper addresses the identified gap above by contributing a mechanistic account to the neural manifold literature. We apply embedding results from the dynamical systems literature to the question; a promising connection to predictive processing is also discussed (Section~\ref{subsec:predictive_processing}).

Specifically, we specialize the general embedding results of \citet{Hart2025} and \citet{Duan2023} to the regular, low-dimensional regime where sensory cortex operates. The specific new results are as follows:
\begin{itemize}[nosep]
\item Proposition~\ref{prop:bunching_regular}: the bunching condition for regular dynamics reduces to a single checkable inequality, $\rho < 1/\kappa$.
\item Proposition~\ref{prop:no_weak_gs}: weak (non-smooth) generalized synchronization does not arise in this regime.
\item Theorem~\ref{thm:main_synthesis}: for regular dynamics, the conditions for a $C^1$ embedding are explicit and mild.
\item Proposition~\ref{prop:prediction_separation}: prediction accuracy constrains representational geometry even when the contraction rate is unknown, with specific perceptual consequences that are developed in Section~\ref{sec:dim_pred_geometry}.
\item Numerical experiments (Section~\ref{subsec:numerical_experiments}): trained $\tanh$ RNNs exhibit the predicted embedding behavior, and reveal that the sufficient conditions are conservative.
\end{itemize}

\section{Preliminaries}
\label{sec:preliminaries}

\subsection{What Drives the Network?}

The network's input comes from observing a physical process: a pendulum swinging, a tone sounding, a head rotating. In biological systems, the observation is made by a sensor (photoreceptors, cochlear hair cells, vestibular organs) whose output feeds into the recurrent circuit we analyze below. We model the physical process as a dynamical system evolving on a bounded, smooth state space $M$. Formally, $M$ is a compact smooth manifold, but circles and tori are the main cases treated in this paper. At each time step the system advances by applying a map $\phi: M \to M$. We require $\phi$ to be smooth and invertible, that is, a $C^r$ diffeomorphism with $r \ge 2$. We write $\Diff^r(M)$ for the set of all such maps.

We work in discrete time throughout. Continuous-time systems reduce to this setting by sampling at a fixed interval ($\phi := \Phi_\tau$ for a continuous flow $\{\Phi_t\}$). In biological circuits the discretization may be performed by the sensor itself: spiking, synaptic release events, or the characteristic integration window of the receiving population.

To make the sensory representation problem mathematically tractable, we analyze a specific subproblem: under what conditions does a driven recurrent network develop an internal manifold that preserves the structure of its sensory input? We approach this through the generalized synchronization framework, which provides sufficient conditions at three successive levels of representational quality. First, under continuous pressure from the input stream, the network's steady-state response defines a continuous map from the states of the driving dynamics to network states (Theorem~\ref{thm:gs_existence}). Second, under a domination condition, this map is smooth (Theorem~\ref{thm:bunching}). Third, when the hidden dimension is large enough and the network and observation function are generic, the map is one-to-one: distinct driving states produce distinct network states (Theorem~\ref{thm:generic_embedding}). These conditions are sufficient but not necessary; their value is that they are explicit and, in the regular regime treated here, quantitatively mild. For readability we assume $\phi, \omega, F$ are $C^2$ throughout, which suffices for all three levels.

\subsection{What Class of Driving Dynamics Do We Study?}
\label{subsec:regular_dynamics}

We define a \emph{regular base system} to be a driving process whose dynamics evolves on a closed curve or higher-dimensional surface that it never leaves, with trajectories that fill it densely without ever exactly repeating.

\begin{definition}[Regular Base Systems]
\label{def:regular_base}
A \emph{regular base system} is a pair $(M, \phi)$ where one of the following holds:
\begin{enumerate}[label=(\roman*)]
\item \textbf{Invariant circle (quasiperiodic):} $M = S^1$ (the unit circle) and $\phi: S^1 \to S^1$ is $C^2$-conjugate (smoothly equivalent, via a twice-differentiable change of coordinates) to an irrational rotation
\[
R_\alpha: \theta \mapsto \theta + \alpha \pmod{1}, \qquad \alpha \in \R \setminus \Q \;\;(\text{$\alpha$ is irrational}).
\]
Because $\alpha$ is not a rational fraction of a full turn, orbits never close and every orbit is dense; the system is \emph{minimal}.

\item \textbf{Invariant torus (quasiperiodic):} $M = T^k := \R^k / \Z^k$ (the $k$-dimensional torus, obtained by identifying opposite faces of the unit cube in $\R^k$) and $\phi: T^k \to T^k$ is $C^2$-conjugate to a translation
\[
T_{\bm{\alpha}}: \bm{\theta} \mapsto \bm{\theta} + \bm{\alpha} \pmod{\Z^k}, \qquad \bm{\alpha} \in \R^k,
\]
with $\{1, \alpha_1, \ldots, \alpha_k\}$ rationally independent (no integer combination of the frequencies equals zero; the oscillations are incommensurate), so every orbit is dense.
\end{enumerate}
\end{definition}

The sensory representation problem motivates the choice of regime. Maps conjugate to rigid rotations arise naturally in the sensory domains discussed in Section~\ref{subsec:sensory_regular}: periodic motion, tonal pitch, rhythm, and head-direction coding all produce effective dynamics of this type. In particular, rigid rotations do not stretch distances, and a conjugate map changes them only through the distortion of the coordinate change. The resulting backward expansion is controlled, so the sufficient conditions require only modest network contraction. Section~\ref{sec:main_results} states the precise inequality. Thus, our analysis reveals that the generalized synchronization framework, under this regime, yields explicit sufficient conditions for a smooth embedding that has desirable properties from the perspective of neural manifold description: the driving dynamics are nearly distortion-free, and the contraction required of the network is correspondingly modest.

As noted, sensors do not observe the full state of the physical world. Section~\ref{subsec:sensory_regular} provides examples where the effective dynamics that the sensor tracks is low-dimensional. The dimension bounds in our theorems depend on the \emph{effective} dimension $d$, not on the complexity of the physical environment. The same reduction keeps the bunching condition mild, because the sensed dynamics separates nearby states less aggressively than the full physical system can.

\subsection{Reconstructing Dynamics from Observations}

A scalar time series can contain enough history to reconstruct the hidden system that generated it. Delay-coordinate embedding (DCE) makes this precise: one collects lagged observations into a vector, and under generic conditions this vector uniquely identifies the system's state.

\begin{definition}[Delay Map]
\label{def:delay_map}
Given $(M, \phi)$, an observation function $\omega: M \to \R$, and embedding dimension $n \in \N$, the delay map is
\[
\Phi_\omega^n: M \to \R^n, \qquad \Phi_\omega^n(x) = \big(\omega(x), \omega(\phi(x)), \ldots, \omega(\phi^{n-1}(x))\big).
\]
\end{definition}

Takens' theorem gives conditions under which the delay map is one-to-one and smooth. The conditions involve the periodic orbits of $\phi$ (trajectories that return to their starting point); in our quasiperiodic setting these conditions are automatically satisfied (Remark~\ref{rem:per_vac}).

\begin{theorem}[Takens {\citep{Takens1981}}]
\label{thm:takens}
Let $M$ be a compact $m$-dimensional manifold and $\phi \in \Diff^2(M)$. Suppose $\phi$ has finitely many periodic orbits, and for each periodic orbit of period $p$, the eigenvalues of $D\phi^p$ are distinct. Then for $n \ge 2m + 1$, a generic (true for most choices in the relevant function space) $\omega \in C^2(M, \R)$ yields a delay map $\Phi_\omega^n$ that is a $C^1$ embedding.
\end{theorem}
The dimension threshold $n \ge 2m+1$ derives from Whitney's embedding theorem \citep{Whitney1936}, which shows that $2m+1$ dimensions suffice to smoothly embed any compact $m$-dimensional manifold. Takens' theorem adapts this sufficient condition to delay maps; the generalized synchronization results of \citet{Hart2025} inherit it in turn.

\begin{remark}[Periodic-orbit hypotheses in quasiperiodic settings]
\label{rem:per_vac}
In the quasiperiodic cases of Definition~\ref{def:regular_base}, $\phi$ has no periodic orbits. Hypotheses in Takens- or Hart-type embedding theorems that impose conditions on periodic orbits are therefore vacuously satisfied (automatically true because no instances exist to violate them).
\end{remark}

For our purposes (smooth manifolds), Takens' theorem suffices. \citet{SauerYorkeCasdagli1991} extend the result to compact sets that need not be manifolds, using box-counting dimension (a measure of the scaling of the number of boxes needed to cover a set) in place of manifold dimension; for a $k$-torus the two coincide, giving $n > 2k$.

\subsection{Recurrent Circuits as Driven Systems}
\label{subsec:driven_systems}

An RNN with hidden state $h_t \in \R^N$ receiving scalar input $u_t$ evolves according to~\eqref{eq:rnn_abstract}. When the input derives from observing a dynamical system ($u_t = \omega(\phi^t(x_0))$), the combined system (environment plus network) evolves jointly. The environment advances by $\phi$; the network updates by $F$ using the current observation. We write this combined update as the skew-product:
\begin{equation}
\label{eq:skew_product}
\Psi: M \times \R^N \to M \times \R^N, \qquad \Psi(x, h) = \big(\phi(x),\; F(h, \omega(x))\big).
\end{equation}

Theorem~\ref{thm:gs_existence} below shows that the network's response settles into a stable relationship with the driving process, so that each driving state produces a unique network state. This occurs under the specific condition that the network forgets its initial state quickly enough. This \emph{echo state property} formalizes this requirement.

\begin{definition}[Echo State Property]
\label{def:esp}
Consider the RNN state update $F$ from~\eqref{eq:rnn_abstract}, with hidden state $h_t \in \R^N$ and scalar input $u_t$. Let $U \subset \R$ be the range of inputs the network receives (a compact set). The RNN satisfies the \emph{echo state property} (ESP) with contraction rate $\rho < 1$ on $U$ if there exists a compact region $K \subset \R^N$ of hidden-state space such that:
\begin{enumerate}[label=(\roman*)]
\item $F(K, u) \subseteq K$ for all $u \in U$ (forward invariance: once the hidden state enters $K$, it stays there).
\item $\sup_{(h,u) \in K \times U} \|\partial_h F(h,u)\|_{\mathrm{op}} \le \rho$, where $\partial_h F$ is the derivative of the state update with respect to the hidden state (measuring how sensitive the next state is to perturbations of the current state) and $\|\cdot\|_{\mathrm{op}}$ is the operator norm (the worst-case multiplicative gain). This says: small perturbations to the hidden state shrink by at least the factor $\rho$ at every step.
\end{enumerate}
\end{definition}

A simple sufficient condition for the echo state property in $\tanh$ recurrent neural networks is $\|W\|_{\mathrm{op}} < 1$, since the $\tanh$ nonlinearity is $1$-Lipschitz and therefore cannot amplify perturbations beyond the gain imposed by the weight matrix \citep{Jaeger2001, LukoseviciusJaeger2009}. Gated architectures such as long short-term memory and gated recurrent unit networks admit related sufficient stability conditions, but these depend on additional architecture-specific inequalities involving the weights and gates \citep{MillerHardt2018,BonassiFarinaScattolini2021}.

Requiring stability for all possible inputs is a strong condition. Biological circuits do not need that. They only need stability for the structured sensory signals they actually receive. Input-dependent versions of ESP formalize that weaker requirement \citep{ManjunathJaeger2013,YildizJaegerKiebel2012}.

Three lines of evidence from sensory cortex are consistent with applying this weaker condition to sensory cortex. First, V1 neurons exhibit fading memory: their responses depend on input from the past $\sim$100--300\,ms but not further back, the signature of a system that forgets old states \citep{NikolicEtAl2009}. Second, stimulus onset quenches trial-to-trial variability in cortical responses, meaning the input forces the network toward a reproducible state regardless of where it started \citep{HennequinEtAl2018}. Third, population responses in visual cortex satisfy a smoothness constraint, in the sense that small changes in input do not dominate population activity \citep{Stringer2019Geom}.

All formal statements in this paper use strict ESP as a sufficient condition. They depend only on the fixed-parameter map $F$ from Definition~\ref{def:rnn} satisfying this property, not on how the parameters $\theta$ are obtained: $F$ may come from a fixed (random) reservoir or from training, and the results apply to the resulting fixed map in either case.

\subsection{When Does a Driven Circuit Settle into a Stable Representation?}

After transients decay, the recurrent circuit settles into a stable correspondence between environmental states and neural states.

\begin{definition}[Generalized Synchronization]
\label{def:gs}
The driven system $\Psi$ (equation~\eqref{eq:skew_product}) exhibits \emph{generalized synchronization} (GS) on $M$ if there exists a continuous function $f: M \to \R^N$ such that:
\begin{enumerate}[label=(\roman*)]
\item \textbf{Invariance:} $f(\phi(x)) = F(f(x), \omega(x))$ for all $x \in M$. (Applying $f$ and then evolving the network gives the same result as evolving the environment and then applying $f$.)
\item \textbf{Attraction:} For any starting environmental state $x_0 \in M$ and any initial hidden state $h_0 \in K$ (the compact region from Definition~\ref{def:esp}), the hidden state $h_t$ produced by iterating~\eqref{eq:rnn_abstract} converges to $f$:
\[
\|h_t - f(\phi^t(x_0))\| \to 0 \quad \text{as } t \to \infty.
\]
\end{enumerate}
The function $f$ is the \emph{synchronization function} or \emph{echo state map}: it sends each environmental state to the hidden state the network converges to when driven from that state.
\end{definition}

\begin{theorem}[GS Existence {\citep{HartHookDawes2020}}]
\label{thm:gs_existence}
Let $(M, \phi)$ be a base system with $\phi \in \Diff^1(M)$, and let $\omega: M \to \R$ be continuous with $U = \omega(M)$ compact. If the RNN satisfies ESP with rate $\rho < 1$ on $U$, then there exists a unique continuous synchronization function $f: M \to \R^N$. Convergence is exponential:
\[
\|h_t - f(\phi^t(x_0))\| = O(\rho^t).
\]
\end{theorem}
\begin{figure}[ht]
\centering
\definecolor{envblue}{HTML}{2166AC}
\definecolor{neuralred}{HTML}{B2182B}
\definecolor{syncgreen}{HTML}{1B7837}
\begin{tikzpicture}[
    node distance=2.8cm and 4.2cm,
    state/.style={rectangle, rounded corners=3pt, draw, minimum width=2cm, minimum height=0.9cm, font=\normalsize},
    env/.style={state, fill=envblue!8, draw=envblue!80!black, line width=0.8pt},
    neural/.style={state, fill=neuralred!8, draw=neuralred!80!black, line width=0.8pt},
    input/.style={font=\small},
    arrow/.style={-{Stealth[length=2.5mm, width=2mm]}, line width=0.9pt},
    emb/.style={-{Stealth[length=2.5mm, width=2mm]}, line width=0.9pt, dashed, envblue!70!black},
    obs/.style={-{Stealth[length=2.2mm, width=1.8mm]}, line width=0.8pt, syncgreen!80!black},
]

\node[env] (xt) {$x_t$};
\node[env, right=of xt] (xt1) {$x_{t+1}$};

\node[neural, below=of xt] (ht) {$h_t$};
\node[neural, below=of xt1] (ht1) {$h_{t+1}$};

\draw[arrow] (xt) -- node[above, font=\small] {$\phi$} (xt1);
\draw[arrow] (ht) -- node[below, font=\small] {$F$} (ht1);

\draw[emb] (xt) -- node[left, font=\small] {$f$} (ht);
\draw[emb] (xt1) -- node[right, font=\small] {$f$} (ht1);

\draw[obs] (xt) to[out=-35, in=145] node[pos=0.38, above right, font=\small, text=syncgreen!70!black] {$u_t=\omega(x_t)$} (ht1);

\node[left=0.9cm of xt, font=\small\itshape, text=envblue!70!black] {Environment};
\node[left=0.9cm of ht, font=\small\itshape, text=neuralred!70!black] {Neural population state};

\end{tikzpicture}

\caption{Generalized synchronization as a commutative diagram. The environment evolves by $\phi$, the neural state evolves by $F$, and the synchronization map $f: M \to \R^N$ sends each environmental state to its corresponding neural state. Commutativity means $f(\phi(x)) = F(f(x), \omega(x))$.}
\label{fig:gs_commutative}
\end{figure}

\subsection{When Does a Stable Representation Become Smooth?}

The echo state property guarantees that the synchronization function is continuous. Within the generalized synchronization framework, upgrading continuity to smoothness requires a stronger condition: the circuit must damp perturbations strongly enough to offset how the driving dynamics separates nearby states when traced backward in time. The required inequality is
\begin{equation}
\label{eq:bunching}
\rho \cdot \|D\phi^{-1}\|_{C^0(M)} < 1,
\end{equation}
where $\rho$ is the ESP contraction rate and
\[
\|D\phi^{-1}\|_{C^0(M)} = \sup_{x \in M} \|D_x\phi^{-1}\|_{\mathrm{op}}
\]
measures the worst-case backward expansion of the driving dynamics. This again is a \emph{bunching condition} in the sense of \citet{HirschPughShub1977}; for skew-product systems, see \citet{stark1999regularity}.

\begin{theorem}[Regularity under Bunching {\citep{stark1999regularity, HartHookDawes2020}}]
\label{thm:bunching}
Under the hypotheses of Theorem~\ref{thm:gs_existence}, suppose additionally that $\phi \in \Diff^2(M)$, $\omega \in C^1(M, \R)$, $F \in C^2(\R^N \times \R, \R^N)$, and the bunching condition~\eqref{eq:bunching} holds. Then the synchronization function $f: M \to \R^N$ is $C^1$.
\end{theorem}

Smoothness alone does not guarantee that the representation preserves state separation. The next theorem gives a sufficient condition for that stronger result.

\begin{theorem}[Generic Embeddings {\citep{Hart2025, Duan2023}}]
\label{thm:generic_embedding}
Let $(M, \phi)$ be a base system with $M$ a compact $m$-dimensional manifold and $\phi \in \Diff^2(M)$ having finitely many periodic orbits, with pairwise distinct eigenvalues at each. In the quasiperiodic cases of Definition~\ref{def:regular_base}, there are no periodic orbits, so these conditions are automatically satisfied. Suppose $N > 2m$ (see discussion on \citet{Whitney1936} in Theorem~\ref{thm:takens}). Then for generic $\omega \in C^1(M, \R)$ and generic state-update maps $F$ (among those satisfying ESP and bunching), the synchronization function $f: M \to \R^N$ is a $C^1$ embedding.
\end{theorem}

Here, \emph{generic} means true outside a residual subset (a countable intersection of open dense sets) of the relevant function space, both for the observation map $\omega\in C^1(M,\R)$ and for the admissible state-update maps $F$. The observation functions and trained networks used in the experiments are specific, not chosen at random from a function space. Section~\ref{subsec:genericity} relates this genericity hypothesis to those specific models: for the observation map the statement strengthens to prevalence, a translation-invariant ``almost every'' on function space, and for the trained networks we certify the predicted failure modes empirically rather than assert genericity.

\subsection{How Do the Mathematical Objects Map onto Biology?}

Table~\ref{tab:levels} links the mathematical objects in the model to components of a biological sensory system. It also shows the two routes by which environmental dynamics can be represented in neural state space: explicit delay coordinates and implicit encoding through recurrent dynamics. Proposition~\ref{prop:finite_lag} shows that the two are closely related.

\begin{table}[ht]
\centering
\renewcommand{\arraystretch}{1.3}
\begin{tabular}{@{}>{\raggedright}p{3.2cm} >{\raggedright}p{4.2cm} p{5.3cm}@{}}
\toprule
\textbf{Level} & \textbf{Mathematical Entity} & \textbf{Neural / Physical Interpretation} \\
\midrule
Environmental dynamics & Manifold $M$, diffeomorphism $\phi$ & Physical process (projectile motion, sound wave, reflectance spectrum) \\[4pt]
Observation function & $\omega: M \to \R$ & Sensory transduction (photoreceptors, cochlear hair cells, mechanoreceptors) \\[4pt]
Input signal & $u_t = \omega(\phi^t(x_0))$ & Afferent activity reaching cortex \\[4pt]
\midrule
\multicolumn{3}{l}{\textit{Two routes to representation:}} \\[2pt]
\midrule
\multicolumn{3}{l}{\uline{Route 1: Explicit DCE}} \\
Delay coordinates & $\Phi_\omega^n: M \to \R^n$ & Explicit storage of input history \\[4pt]
\midrule
\multicolumn{3}{l}{\uline{Route 2: Generalized synchronization}} \\
Hidden state & $h_t \in \R^N$ & Mesoscale population activity \\[4pt]
Recurrent dynamics & $F: \R^N \times \R \to \R^N$ & Recurrent cortical dynamics \\[4pt]
Synchronization function & $f: M \to \R^N$ & Learned representation shaped by prediction \\[4pt]
\bottomrule
\end{tabular}

\smallskip
\noindent\textit{Connection (Proposition~\ref{prop:finite_lag}):} The synchronization function $f$ is approximately a function of a delay vector:
\[
f(x) \approx g_L(\Phi_\omega^L(\phi^{-L}(x))).
\]

\caption{Model components and their biological interpretation. Environmental dynamics can be represented either explicitly through delay coordinates or implicitly through recurrent dynamics.}
\label{tab:levels}
\end{table}

\section{Main Results}
\label{sec:main_results}

How strongly must a sensory circuit damp its own activity to track a moving stimulus faithfully? Biologically this is the question of how fast the circuit forgets past input, the fading memory measured in cortex; mathematically it is set by how aggressively the sensed dynamics separates nearby states. We make it quantitative for the regular regime. Section~\ref{sec:preliminaries} introduced the general results needed to describe when a driven recurrent circuit develops a stable representation of its input. We now apply those results to the regular, low-dimensional regimes. In these regimes, backward expansion is controlled and the smoothness requirement becomes easy to check.

\subsection{How Strong Must Contraction Be?}

For rigid rotations, backward expansion is exactly 1, so any contraction rate $\rho < 1$ suffices. For maps that are only conjugate to rigid rotations, backward expansion is controlled by the distortion of the conjugacy. The next proposition makes this explicit.

\begin{proposition}[Near-Isometric Bounds for Regular Bases]
\label{prop:bunching_regular}
Let $(M, \phi)$ be a regular base system (Definition~\ref{def:regular_base}).
\begin{enumerate}[label=(\roman*)]
\item If $\phi$ is a rigid rotation on $S^1$ or rigid translation on $T^k$, then $\|D\phi^{-1}\|_{C^0(M)} = 1$.
\item If $\phi$ is $C^2$-conjugate to a rigid rotation/translation via a diffeomorphism $h: M \to M$, then
\[
\|D\phi^{-1}\|_{C^0(M)} \le \|Dh\|_{C^0(M)} \, \|Dh^{-1}\|_{C^0(M)}.
\]
\end{enumerate}
Define the conjugacy condition number
\[
\kappa := \|Dh\|_{C^0(M)} \, \|Dh^{-1}\|_{C^0(M)}.
\]
This measures how far the coordinate change is from being distance-preserving; $\kappa = 1$ means no distortion. The bunching condition then becomes $\rho < 1/\kappa$.
\end{proposition}

\begin{proof}
(i) Rigid rotation $R_\alpha: \theta \mapsto \theta + \alpha$ on $S^1$ has $DR_\alpha = 1$ everywhere. Thus $DR_\alpha^{-1} = 1$ and $\|D\phi^{-1}\|_{C^0} = 1$. The argument for translations on $T^k$ is identical.

(ii) Write $\phi = h \circ R_\alpha \circ h^{-1}$, so $\phi^{-1} = h \circ R_\alpha^{-1} \circ h^{-1}$. By the chain rule, with each Jacobian evaluated at the basepoint indicated by the bar,
\[
D_x\phi^{-1} = (Dh)\big|_{\,R_\alpha^{-1}(h^{-1}(x))}\;\cdot\;(DR_\alpha^{-1})\big|_{\,h^{-1}(x)}\;\cdot\;(Dh^{-1})\big|_{\,x}.
\]
Since $DR_\alpha^{-1} = \mathrm{Id}$, this simplifies to $D_x\phi^{-1} = (Dh)\big|_{\,R_\alpha^{-1}(h^{-1}(x))}\,(Dh^{-1})\big|_{\,x}$. Taking operator norms and suprema over $x \in M$ yields the claimed bound.
\end{proof}

Proposition~\ref{prop:bunching_regular} makes the smoothness condition explicit in the regular regime. We now use that bound to show that weak generalized synchronization does not arise there.

\subsection{Bunching Yields $C^1$ Synchronization in the Regular Regime}

When backward expansion dominates contraction, the synchronization function can be continuous but not differentiable. This is weak generalized synchronization \citep{KellerJafriRamaswamy2013}. In the regular regime, the bound above rules it out.

\begin{proposition}[Bunching Excludes Weak GS for Regular Bases]
\label{prop:no_weak_gs}
For a regular base system with $\|D\phi^{-1}\|_{C^0(M)} \le \kappa$, if the RNN satisfies ESP with rate $\rho < 1/\kappa$, then the synchronization function is $C^1$. Weak (non-smooth) generalized synchronization does not arise.
\end{proposition}

\begin{proof}
Proposition~\ref{prop:bunching_regular} gives $\rho \cdot \|D\phi^{-1}\|_{C^0(M)} \le \rho \kappa < 1$, so the bunching condition holds. Theorem~\ref{thm:bunching} then guarantees $C^1$ regularity. 
\end{proof}
The $C^1$ regularity is not imposed as a requirement; it arises as a consequence of the contraction and near-isometry conditions that hold in this regime.

\subsection{Main Theorem}

The previous results did two things. Proposition~\ref{prop:bunching_regular} made the smoothness condition explicit in the regular regime. Proposition~\ref{prop:no_weak_gs} showed that this regime excludes non-smooth synchronization maps. 

We now combine the bunching bound with the theorems from Section~\ref{sec:preliminaries}. For convenience, we collect the assumptions in a single block.

\begin{assumption}[Standard Setup]
\label{assump:standard}
Throughout this section:
\begin{enumerate}[label=(A\arabic*)]
\item $(M, \phi)$ is a regular base system (Definition~\ref{def:regular_base}) with $M$ a $d$-dimensional manifold ($d = 1$ for the circle, $d = k$ for the $k$-torus).
\item $\phi \in \Diff^2(M)$ and there exists a constant $\kappa \ge 1$ such that
\[
\|D\phi^{-1}\|_{C^0(M)} := \sup_{x \in M} \|D_x\phi^{-1}\|_{\mathrm{op}} \le \kappa.
\]
\item The observation function $\omega: M \to \R$ is $C^2$.
\item The RNN $F: \R^N \times \R \to \R^N$ is $C^2$ and satisfies ESP with rate $\rho < 1/\kappa$.
\item $N > 2d$.
\end{enumerate}
\end{assumption}

\begin{theorem}[Embedding for Regular Base Systems]
\label{thm:main_synthesis}
Under Assumption~\ref{assump:standard}:
\begin{enumerate}[label=(\roman*)]
\item There exists a unique $C^1$ synchronization function $f: M \to \R^N$.
\item For generic $\omega$ and generic $F$ (among those satisfying (A4)), the map $f$ is a $C^1$ embedding.
\item If $f$ is injective, then the dynamics on $f(M)$ are conjugate to $\phi$ via
\[
\psi := f \circ \phi \circ f^{-1}: f(M) \to f(M),
\]
so that $\psi \circ f = f \circ \phi$. If $f$ is a $C^1$ embedding, then $\psi$ is a $C^1$ diffeomorphism of the embedded submanifold $f(M)$. By invariance, $\psi(z) = F(z, \omega(f^{-1}(z)))$ for all $z \in f(M)$.
\end{enumerate}
\end{theorem}

\begin{proof}
(i) Assumption~(A4) gives ESP on $U = \omega(M)$, so Theorem~\ref{thm:gs_existence} yields a unique continuous synchronization function $f: M \to \R^N$ satisfying $f(\phi(x)) = F(f(x), \omega(x))$ for all $x \in M$. Assumptions~(A2) and (A4) give $\rho \cdot \|D\phi^{-1}\|_{C^0(M)} \le \rho\kappa < 1$, so the bunching condition~\eqref{eq:bunching} holds and Theorem~\ref{thm:bunching} upgrades $f$ to $C^1$.

(ii) In the quasiperiodic cases of Definition~\ref{def:regular_base}, $\phi$ has no periodic points, so the periodic-orbit hypotheses in Theorem~\ref{thm:generic_embedding} are automatically satisfied (Remark~\ref{rem:per_vac}). With $N > 2d$ from Assumption~(A5), Theorem~\ref{thm:generic_embedding} applies.

(iii) Suppose $f$ is injective. Since $M$ is compact and $f(M) \subset \R^N$ is Hausdorff, the continuous bijection $f: M \to f(M)$ is a homeomorphism. Define $\psi := f \circ \phi \circ f^{-1}$. For $z \in f(M)$, writing $z = f(x)$ with $x = f^{-1}(z)$, the invariance relation gives
\[
\psi(z) = f(\phi(x)) = F(f(x), \omega(x)) = F(z, \omega(f^{-1}(z))).
\]
If $f$ is a $C^1$ embedding, then $f^{-1}$ is $C^1$ and $\psi$ is a $C^1$ diffeomorphism of $f(M)$.
\end{proof}

\subsection{Genericity, Prevalence, and Empirical Certification}
\label{subsec:genericity}
Theorem~\ref{thm:main_synthesis}(ii) gives a $C^1$ embedding for generic observation functions and generic state-update maps. Since the numerical experiments below use specific observation functions, such as cosine tuning curves and sine sums, together with specific trained networks, it is important to clarify the role of these genericity assumptions. The theorem should not be read as certifying any particular trained network. Rather, it identifies a large admissible class in which embedding failure is nongeneric, and the experiments then test directly for the finite-sample signatures of such failure.

For the observation function, the appropriate notion of largeness is not ordinary Lebesgue measure. Infinite-dimensional function spaces do not carry a canonical Lebesgue measure, so \emph{almost every} must be replaced by a measure-theoretic analogue. A property is called \emph{prevalent} if it holds outside a \emph{shy} set, the infinite-dimensional analogue of a measure-zero set \citep{HuntSauerYorke1992}. In the delay-coordinate setting of Takens' theorem, \citet{SauerYorkeCasdagli1991} showed that, above the relevant dimension threshold, the set of observation functions for which the delay map fails to be injective is shy. Thus injectivity is prevalent in the observation class. This is a statement about the function class as a whole, not a pointwise certificate for a prescribed observation function. Consequently, the specific cosine and sine observation functions used below are not certified merely by prevalence.

A similar distinction applies to the reservoir or recurrent state-update map. Once the required contraction hypotheses hold, existence and regularity of the synchronization map follow from invariant-graph regularity results such as those of \citet{stark1999regularity}. The embedding conclusion then requires the additional genericity, dimension, and regularity hypotheses appearing in the reservoir embedding theorems of \citet{Hart2025} and \citet{Duan2023}. These results apply to a fixed state-update map only when that map lies in the generic admissible class; the map need not have been generated randomly. After training, a $\tanh$ network defines a fixed map $F_\theta$. If this trained map satisfies the echo-state, bunching, regularity, dimension, and genericity hypotheses, the same embedding conclusion applies. No separate theorem for \emph{trained} networks is required.

What the theorem does not establish is that gradient descent necessarily selects an admissible generic state-update map or preserves strict contraction throughout training. For this reason, the numerical experiments are not presented as formal certifications. Instead, they test the trained networks for the concrete failure modes that nongenericity or loss of contraction would produce: failure of initial conditions to synchronize (Experiment~1), approximate collisions in which well-separated sensory states map to nearly coincident hidden states (Experiment~2), low effective rank (Experiment~3), and failure of topology recovery (Experiment~4). These tests are finite-sample diagnostics rather than proofs. Nevertheless, they are aligned with the theory: a synchronization map that failed to embed would show approximate collisions that do not vanish as the hidden dimension grows, whereas the measured collision fractions instead fall toward zero once the hidden dimension passes the theoretical threshold and stay low across the three random seeds (Section~\ref{subsec:numerical_experiments}).

Finally, the distinct-future separation implied by injectivity is the content of Lemma~\ref{lem:gamma_injective}, not of the collision diagnostic itself. The graceful sub-threshold degradation described by \citet{SauerYorkeCasdagli1991} concerns the regime $N<2d+1$; it is distinct from whether a given observation function lies in the shy failure set at fixed above-threshold $N$.
\par

\subsection{Uniqueness of the Synchronization Function}

Theorem~\ref{thm:main_synthesis} guarantees that the synchronization function exists. A separate question is whether it is unique. This matters because if multiple synchronization functions were compatible with the same dynamics, the representation would be underdetermined and the embedding would lack a definite geometric interpretation. It also matters for a reason that will be developed further in Section~\ref{sec:discussion}: whether the representation is determined by the full temporal process or by any finite collection of input-output snapshots.

\begin{proposition}[Uniqueness]
\label{prop:processual_uniqueness}
Under ESP with rate $\rho < 1$:
\begin{enumerate}[label=(\roman*)]
\item The invariance condition $f(\phi(x)) = F(f(x), \omega(x))$ for all $x \in M$ admits exactly one continuous solution $f: M \to \R^N$.
\item For any finite subset $S = \{x_1, \ldots, x_n\} \subset M$, the pointwise constraints $g(x_i) = f(x_i)$ for $i = 1, \ldots, n$ do not uniquely determine $g$ among continuous maps $M \to \R^N$.
\end{enumerate}
\end{proposition}

\begin{proof}
Part (i) is Theorem~\ref{thm:gs_existence}. For part (ii), since $M$ is a positive-dimensional compact manifold, $M \setminus S$ is nonempty and open. Choose any $x^* \in M \setminus S$. A continuous bump function supported in a neighborhood of $x^*$ disjoint from $S$ gives a map $g: M \to \R^N$ with $g(x_i) = f(x_i)$ for all $i$ but $g(x^*) \neq f(x^*)$.
\end{proof}

This proposition isolates the role of the invariance condition: finite pointwise matches do not determine the synchronization map, but the full dynamical constraint does.

\section{Dimensions, Prediction, and Representational Geometry}
\label{sec:dim_pred_geometry}

Two questions a sensory circuit must answer have direct neural counterparts: how many neurons does a faithful representation need, and how finely can the circuit tell stimuli apart? We address the first through the dimension of the embedding and the second through prediction accuracy, which together set the resolution of the representation. The main theorem (Section~\ref{sec:main_results}) gives sufficient conditions for a smooth embedding. This section develops the consequences: how much neural state space the embedding requires, how it connects to delay-coordinate reconstruction, and what happens when contraction properties are unknown but prediction accuracy is measurable.

\subsection{Relation to Delay Coordinates}

Theorem~\ref{thm:main_synthesis} is proved through generalized synchronization and does not require Takens' theorem. The connection to delay coordinates matters for two reasons.

First, delay-coordinate embedding is the reconstruction method most readers will know. The next proposition shows that the synchronization function $f$ approximately encodes a delay vector: the network's hidden state carries the same information as a finite window of past observations, with the approximation improving exponentially as the window grows.

Second, the connection becomes essential below. The prediction-separation result (Proposition~\ref{prop:prediction_separation}) requires the forward observation map $\Gamma_K$ to be one-to-one for large enough $K$. That injectivity comes from Takens' theorem.

\begin{proposition}[Finite-Lag Approximation {\citep{HartHookDawes2020}}]
\label{prop:finite_lag}
Under Assumption~\ref{assump:standard}, for any $\varepsilon > 0$ there exists $L \in \N$ and a continuous map $g_L: \R^L \to \R^N$ such that
\[
\sup_{x \in M} \|f(x) - g_L(\omega(\phi^{-L}(x)), \ldots, \omega(\phi^{-1}(x)))\| < \varepsilon.
\]
The error decays as $O(\rho^L)$.
\end{proposition}

\subsection{Dimension Bounds}
\label{sec:dimensions}

Theorem~\ref{thm:main_synthesis} has a sufficient condition of $N > 2d$, where $d$ is the dimension of the sensory dynamics being tracked rather than the full physical environment. This condition, as noted in Theorem~\ref{thm:takens}, is inherited from Whitney's embedding theorem \citep{Whitney1936}. For circles and low-dimensional tori, this yields modest dimension conditions for embedding (which holds for generic networks and observation functions): $N \ge 3$ for $d=1$, $N \ge 5$ for $d=2$, and in general $N \ge 2d+1$. The bound scales with intrinsic sensory structure, not with the complexity of the world.

The $N > 2d$ condition is sufficient but not strictly necessary. For delay-coordinate maps, \citet{SauerYorkeCasdagli1991} showed that when $N \le 2d$, the set on which one-to-one fails (the self-intersection set) has box-counting dimension at most $2d - N$. For $d = 1$ and $N = 2$, self-intersections can occur, but only on a set of dimension at most zero (a discrete set of points). Whether this graceful-degradation property transfers from delay maps to synchronization maps is not proved here, but the pattern is consistent with Experiment~2 (Section~\ref{subsec:numerical_experiments}), where collision fractions are small but nonzero below the threshold and decrease smoothly as $N$ increases, rather than vanishing at a sharp boundary.

This scale is biologically plausible. Empirical studies of neural population dynamics report low-dimensional but still substantial state spaces: motor cortex activity during reaching lies on manifolds of dimension about 10--15 \citep{ChurchlandEtAl2012}, visual cortex responses to natural images are captured by projections of dimension about 10--20 \citep{StringerEtAl2019}, and prefrontal activity during working memory also evolves on low-dimensional manifolds \citep{MurrayEtAl2017}. These values exceed the requirements for regular sensory dynamics of the kind considered here. Table~\ref{tab:perception_dimensions} summarizes the dimensional requirements for dynamical settings motivated by sensory processing.

\begin{table}[ht]
\centering
\renewcommand{\arraystretch}{1.2}
\begin{tabular}{@{}>{\raggedright}p{4cm} >{\raggedright}p{4cm} cc@{}}
\toprule
\textbf{Dynamical Setting} & \textbf{State Space Geometry} & \textbf{Intrinsic $d$} & \textbf{Sufficient $N$} \\
\midrule
Simple periodic motion / rotation & Circle ($S^1$) & 1 & $3$ \\
Gait with a dominant cycle & Limit cycle (intrinsic $S^1$) & 1 & $3$ \\
Pure tone (pitch) & Circle ($S^1$) & 1 & $3$ \\
Complex tone / chords & Torus ($T^k$, $k \le 5$) & $\le 5$ & $11$ \\
Simple rhythm & Circle ($S^1$) & 1 & $3$ \\
Polyrhythm & Torus ($T^k$, $k \le 3$) & $\le 3$ & $7$ \\
Color opponent space (slow adaptation) & Slow manifold in $\R^3$ & $\le 3$ & $7$ \\
\bottomrule
\end{tabular}
\caption{Dimension requirements for embedding dynamical settings arising in sensory processing. ``Intrinsic $d$'' is the dimension of the driving dynamics; ``Sufficient $N$'' is the minimum hidden dimension for a smooth embedding ($N > 2d$, for generic observation and network). The color opponent-space row is motivational: opponent processing uses three channels (one luminance, two chromatic), giving $d \le 3$, but slow chromatic drift need not satisfy Definition~\ref{def:regular_base} (Section~\ref{subsec:sensory_regular}).}
\label{tab:perception_dimensions}
\end{table}

\subsection{Prediction Accuracy Constrains Representation}

Even when contraction properties are unknown, prediction accuracy constrains the form of the learned representation. If a network predicts future observations accurately, distinct underlying states cannot collapse arbitrarily onto the same internal state. For long enough horizons, the sequence of future observations distinguishes states of the underlying dynamics. Any predictor that reconstructs those futures from the neural state must preserve some degree of state separation.

\subsubsection{Future Observations Distinguish States}

\begin{definition}[$K$-Step Forward Map]
For $K \in \N$, define
\[
\Gamma_K: M \to \R^{K+1}, \qquad \Gamma_K(x) = \big(\omega(x), \omega(\phi(x)), \ldots, \omega(\phi^K(x))\big).
\]
\end{definition}

\begin{lemma}
\label{lem:gamma_injective}
Under the hypotheses of Takens' theorem, for $K \ge 2d$ and generic $\omega$, the map $\Gamma_K$ is injective on $M$.
\end{lemma}

\begin{proof}[Proof sketch]
The result is a direct application of the delay-embedding theorem; we record the steps. The map $\Gamma_K$ is the delay map of Definition~\ref{def:delay_map} with $n = K+1$, since both read the observation $\omega$ along the forward iterates $\omega(\phi^k(x))$ for $k = 0,\dots,K$. By Theorem~\ref{thm:takens}, for $n = K+1 \ge 2d+1$ (equivalently $K \ge 2d$) and generic $\omega$, the delay map $\Phi_\omega^{K+1}$ is a $C^1$ embedding, hence injective on $M$; the genericity holds in the stronger prevalence sense of \citet{SauerYorkeCasdagli1991} (Section~\ref{subsec:genericity}). Because $\phi$ is a diffeomorphism (Definition~\ref{def:regular_base}), $\Gamma_K$ is the forward delay-coordinate map of the diffeomorphism embedding theorem \citep[Theorem~2.7]{SauerYorkeCasdagli1991}. The classical past-delay reconstruction taken at $\phi^K(x)$ is the coordinate reversal of $\Gamma_K(x)$; since $x \mapsto \phi^K(x)$ is a bijection of $M$, the forward map $\Gamma_K$ and the past-delay map are injective on $M$ together.
\end{proof}

\subsubsection{Prediction Implies State Separation}

Suppose a predictor $P: \R^N \to \R^{K+1}$ approximates the future-observation map on the synchronization image, so that $P(f(x)) \approx \Gamma_K(x)$ for $x \in M$. To quantify how prediction error limits collapse in the representation, we use a separation modulus.

\begin{definition}[Separation modulus]
Let $g: M \to \R^p$ be continuous on compact $M$, and let $d_M$ denote distance on $M$. A \emph{separation modulus} for $g$ is any nondecreasing function $\eta: (0,\operatorname{diam}(M)] \to (0,\infty)$ such that
\[
d_M(x,x') \ge \delta \implies \|g(x)-g(x')\| \ge \eta(\delta).
\]
Any continuous injective map on compact $M$ admits a positive separation modulus. The proposition below turns that intuition into a quantitative bound.
\end{definition}

\begin{proposition}[Prediction--separation link]
\label{prop:prediction_separation}
Let $f: M \to \R^N$ be the synchronization map. Fix $K \in \N$ and define $\Gamma_K$ as above. Assume $\Gamma_K$ has separation modulus $\eta$, and let $P: \R^N \to \R^{K+1}$ satisfy
\[
\sup_{x \in M} \|P(f(x)) - \Gamma_K(x)\| \le \varepsilon .
\]
Then for every $\delta > 0$ such that $2\varepsilon < \eta(\delta)$,
\[
f(x)=f(x') \implies d_M(x,x') < \delta .
\]
\end{proposition}

\begin{proof}
Assume $f(x)=f(x')$. Then $P(f(x))=P(f(x'))$. By the uniform error bound,
\[
\|\Gamma_K(x)-\Gamma_K(x')\|
\le
\|\Gamma_K(x)-P(f(x))\|+\|P(f(x'))-\Gamma_K(x')\|
\le 2\varepsilon .
\]
If $d_M(x,x') \ge \delta$, the separation modulus gives
\[
\|\Gamma_K(x)-\Gamma_K(x')\| \ge \eta(\delta) > 2\varepsilon ,
\]
a contradiction. Therefore $d_M(x,x') < \delta$.
\end{proof}

When $2\varepsilon \ge \sup_\delta \eta(\delta)$, the proposition yields no separation guarantee at any prescribed scale.

\begin{corollary}[Quantitative separation]
\label{cor:quantitative_separation}
If in addition $P$ is $L$-Lipschitz on $f(M)$ for some $L>0$, then
\[
\|f(x)-f(x')\| \ge \frac{\max\{0,\|\Gamma_K(x)-\Gamma_K(x')\|-2\varepsilon\}}{L}.
\]
\end{corollary}

Proposition~\ref{prop:prediction_separation} assumes a uniform error bound, while training controls expected error. The next corollary replaces the uniform bound by one that holds with high probability.

\begin{corollary}[Probabilistic separation]
\label{cor:probabilistic_separation}
Suppose the uniform bound is relaxed to hold on a set $G \subseteq M$ of measure $\mu(G) \ge 1-p$, that is $\|P(f(x))-\Gamma_K(x)\| \le \varepsilon$ for $x \in G$. If $x$ and $x'$ are drawn independently from the invariant measure $\mu$, the separation conclusion of Proposition~\ref{prop:prediction_separation} holds for the pair $(x,x')$ with probability at least $(1-p)^2$, so it can fail on at most a fraction $1-(1-p)^2 = 2p-p^2 \le 2p$ of pairs.
\end{corollary}
\begin{proof}
The argument is elementary. The proof of Proposition~\ref{prop:prediction_separation} invokes the error bound only at the two points $x$ and $x'$, so its conclusion holds whenever both lie in $G$. Under independent sampling from $\mu$, $\Pr(x \in G \text{ and } x' \in G) = \mu(G)^2 \ge (1-p)^2$, so the guarantee fails for at most a fraction $1-(1-p)^2 = 2p-p^2 \le 2p$ of pairs.
\end{proof}
\subsubsection{Perceptual Consequences}

The prediction-separation link does not give a perfect embedding. It gives a scale-limited one. Three factors limit the guarantee, and each corresponds to a familiar feature of perception. 
First, training minimizes expected prediction error, not uniform error. Distinctions that the organism encounters frequently will be predicted well and therefore represented finely. Rare distinctions will be predicted poorly and may collapse. This is categorical perception: a continuum of physical stimuli perceived as falling into discrete categories, with better discrimination across category boundaries than within them \citep{Harnad1987}. The same logic predicts expertise effects. Musicians show enhanced pitch discrimination relative to non-musicians \citep{Micheyl2006, Bidelman2011}. Similarly, trained oenologists make fine-grained discriminations among wines that novices perceive as equivalent \citep{Smith2007-oe}, an expertise effect that \citet{Dolega2025-bl} situate within the quality space framework. In both, learning reshapes the geometry of the representational space, increasing separation along task-relevant dimensions. In the present framework, this would correspond to finer state separation in the relevant hidden manifold, driven by lower prediction error for frequently encountered stimuli.

Second, the separation modulus $\eta$ of the forward observation map $\Gamma_K$ varies across the driving manifold. States whose $K$-step futures are nearly identical produce small $\eta$, regardless of how far apart they are on the base manifold. The representation can therefore collapse physically distinct states that happen to generate the same future observations. In color vision, different spectral distributions that produce indistinguishable perceptual responses are called metamers. The present framework generalizes this: metameric collapse occurs whenever distinct driving states have near-identical futures, whether the states are colors, pitches, or spatial orientations.

Third, finite prediction error $\varepsilon$ sets a hard resolution limit. Proposition~\ref{prop:prediction_separation} guarantees separation only for states whose futures differ by more than $2\varepsilon$. Below that scale, distinct states may be indistinguishable in the representation. This is a discrimination threshold: the minimum physical difference that the system can resolve.

\begin{table}[ht]
\centering
\renewcommand{\arraystretch}{1.2}
\begin{tabular}{@{}>{\raggedright}p{5.5cm} p{6.5cm}@{}}
\toprule
\textbf{Limitation of the guarantee} & \textbf{Perceptual consequence} \\
\midrule
Uniform vs.\ expected prediction error & Categorical perception, expertise effects \\
Variation in separation modulus $\eta$ & Metameric collapse \\
Finite resolution set by $\varepsilon$ & Discrimination thresholds \\
\bottomrule
\end{tabular}
\caption{Limits of the prediction--separation guarantee and their perceptual consequences.}
\label{tab:gaps}
\end{table}

\section{Numerical Illustrations}
\label{sec:numerical}

\subsection{Sensory Dynamics as Regular Structure}
\label{subsec:sensory_regular}
The theorems above apply to driving dynamics on circles and tori. This subsection identifies sensory domains where such dynamics arise, grounding the mathematical regime in perception.

Motion tracking is the simplest case. Pendular motion traces a limit cycle, a closed and repeating trajectory in state space, and steady rotation is dynamics on a circle; coordinated gait and gesture are well approximated by low-dimensional oscillatory structure \citep{ZagoEtAl2009, Kelso1995, VyasEtAl2020}.

Auditory processing supplies the torus. A pure tone is a limit cycle in phase space, and a chord with $k$ incommensurate frequencies (frequencies whose ratios are irrational) is quasiperiodic motion on a $k$-dimensional torus. Speech adds sequences of relatively stable configurations, formant patterns for vowels and transient trajectories for consonants \citep{SaltzmanMunhall1989, Port2003}, while rhythm contributes periodic structure \citep{LargeKolen1994}.

Color vision is a motivating case that lies outside the formal scope. The visual system encodes color as differences between cone responses rather than as raw cone intensities, an opponent-process representation that is low-dimensional \citep{Wandell1995}; color-constancy mechanisms stabilize perception across illumination changes, and the relevant dynamics is a slow drift on a low-dimensional manifold \citep{BrainardFreeman1997}. That drift is not in general a minimal quasiperiodic flow on a torus, so it need not satisfy Definition~\ref{def:regular_base}: applying the embedding results here would require restricting to a regime conjugate to a rotation, or extending them to general low-dimensional attracting manifolds (Section~\ref{subsec:limitations}).

Head-direction coding spans both shapes. Azimuthal head direction is a ring ($S^1$), and head-direction cells in \emph{Drosophila} and rodents form ring-shaped manifolds \citep{Kim2017, Chaudhuri2019}. Three-dimensional head direction adds a periodic angle: in the Egyptian fruit bat, azimuth and pitch are coded together in toroidal coordinates \citep{Finkelstein2015}, and grid cell populations in rodents, a periodic spatial code, exhibit toroidal organization \citep{Gardner2022}. These cases illustrate topology-preserving but metrically imperfect representation: the shape of the manifold is correct, but distances on it are not uniformly faithful.

In each domain, the sensory apparatus filters the physical world into a signal whose effective dynamics is low-dimensional and regular. The theorems in Sections~\ref{sec:main_results}--\ref{sec:dim_pred_geometry} characterize when a recurrent circuit receiving such a signal builds a faithful internal model of it.
\par

\subsection{Worked Example}
\label{subsec:numerical_example}

Theorem~\ref{thm:main_synthesis} gives explicit sufficient conditions. This example shows that, in a simple sensory setting, those conditions are easy to satisfy and require only a modest hidden dimension.

Consider a rigid translation on $T^2 = \R^2/\Z^2$ with rationally independent frequencies $\alpha_1 = 1/\sqrt{2}$ and $\alpha_2 = 1/\sqrt{3}$. This is the kind of quasiperiodic signal produced by a two-frequency sensory drive. In the standard flat metric, $D\phi \equiv I$, so $\kappa = 1$. A $\tanh$ recurrent neural network~\eqref{eq:rnn_tanh} with $W = sQ$, where $Q$ is orthogonal and $0 < s < 1$, satisfies
\[
\|\partial_h F\|_{\mathrm{op}} \le \|W\|_{\mathrm{op}} = s.
\]
Choosing $s = 0.95$ gives $\rho \le 0.95$, so the bunching condition holds:
\[
\rho \cdot \|D\phi^{-1}\|_{C^0} \le 0.95 < 1.
\]
In the rigid case, strict ESP is enough.

For $C^2$-conjugate driving with conjugacy condition number $\kappa = 1.05$, bunching requires
\[
\rho < 1/1.05 \approx 0.952,
\]
which is still satisfied by $\rho = 0.95$. Near-isometric driving therefore keeps the smoothness condition mild. Since $d = 2$, Theorem~\ref{thm:main_synthesis} requires $N > 4$, that is, $N \ge 5$.

The same pattern holds more generally. For sensory dynamics on a circle ($d = 1$) or a low-dimensional torus ($d = 2$ or $3$), the dimension requirement is modest ($N \ge 3$, $5$, or $7$), and the contraction requirement is close to the minimum needed for ESP. The conditions become demanding only when the conjugacy distortion $\kappa$ is large or the torus dimension $k$ is high. Izhikevich-type bursting neurons provide another example: their dynamics lie on a slow manifold well approximated by a limit cycle on $S^1$ \citep{Izhikevich2003, Izhikevich2006}, placing them in the $d = 1$, $N \ge 3$ case. Section~\ref{sec:dimensions} summarizes the corresponding sensory cases.

\subsection{Numerical Experiments}
\label{subsec:numerical_experiments}

Training relates to Sections~\ref{sec:preliminaries}--\ref{sec:dimensions} through a separation of timescales between the parameters $\theta$ and the hidden state $h_t$. The parameters are structural: fixed within a trial, changed only slowly by learning, or not changed at all. The hidden state is the fast variable, recomputed at every step. So the evolution of $h_t$ is not the learning: $h_t$ is the fast inference variable, and the parameters carry whatever learning has occurred. The theorems of Sections~\ref{sec:preliminaries}--\ref{sec:dimensions} concern the fixed-parameter map $F$, not how its parameters were obtained, so they apply to a learned $F$ exactly as to a fixed one whenever that $F$ satisfies their hypotheses. The synchronization function $f:M\to\R^N$ of Section~\ref{sec:preliminaries} sends each environmental state to the hidden state the network settles into when driven by it, and it is a property of $F$. Three results from those sections set its quality. Existence and continuity of $f$ follow from contraction of $F$ in the hidden state, the echo state property (Theorem~\ref{thm:gs_existence} in Section~\ref{sec:preliminaries}). Smoothness follows from that contraction being strong relative to how the drive separates nearby states traced backward in time, the bunching condition (Theorem~\ref{thm:bunching}; the regular-regime bound that verifies it is Proposition~\ref{prop:bunching_regular} in Section~\ref{sec:main_results}). Injectivity follows from the hidden dimension exceeding twice the sensory dimension, $N>2d$, with genericity of the observation function and the network (Theorem~\ref{thm:main_synthesis} in Section~\ref{sec:main_results}, with the dimension bounds developed in Section~\ref{sec:dimensions}). Training enters only by producing the particular $F$ that these results are then applied to.

Each experiment instantiates a concrete sensory scenario from Section~\ref{subsec:sensory_regular}. A head-direction-like signal on a circle $S^1$ drives Experiments~1 and~4, and a two-frequency, chord-like signal on a torus $T^2$ drives Experiments~2 and~4; Experiment~3 studies prediction in a head-direction-like circuit, and Experiment~1 additionally includes a chaotic (logistic-map) driver as a non-biological contrast to the regular regime. The mathematical objects (the synchronization map, its dimension, its separation modulus) are the model's account of how such circuits represent these signals.

Four numerical experiments test the theoretical predictions. All use $\tanh$ recurrent neural networks (equation~\eqref{eq:rnn_tanh}) with hidden state $h_t \in \R^N$, trained by truncated backpropagation through time on a one-step prediction objective. The recurrent weight matrix $W$ is initialized with operator norm $\|W\|_{\mathrm{op}} = s$ for a prescribed spectral scale $s < 1$ (the operator norm is the largest singular value of $W$, the gain controlling the echo state property in Definition~\ref{def:esp}; the initialization procedure is detailed in Appendix~\ref{app:experiments}). After training, we verify whether contraction is maintained by recomputing $\|W\|_{\mathrm{op}}$ and the spectral radius. We train the recurrent weights, rather than fixing a reservoir, because the biological question concerns circuits that adapt their connectivity; the post-training operator norm can exceed $1$ as a result (Experiment~1).

\subsubsection{Experiment 1: Generalized Synchronization Under Regular and Chaotic Driving}

This experiment illustrates the role of driving regularity and contraction strength in generalized synchronization. Two drivers (quasiperiodic rotation on $S^1$, where Proposition~\ref{prop:bunching_regular} applies, versus the chaotic logistic map at $r = 3.9$, which falls outside the regular regime) are crossed with two contraction strengths (initial $s = 0.5$ versus $s = 0.95$). The chaotic driver serves as a contrast case. All four conditions use $N = 8$, trained for 200 epochs. Generalized synchronization is assessed by driving the trained network with the same input sequence from five different random initial hidden states and measuring whether the trajectories converge.

\begin{figure}[ht]
\centering
\includegraphics[width=0.49\textwidth]{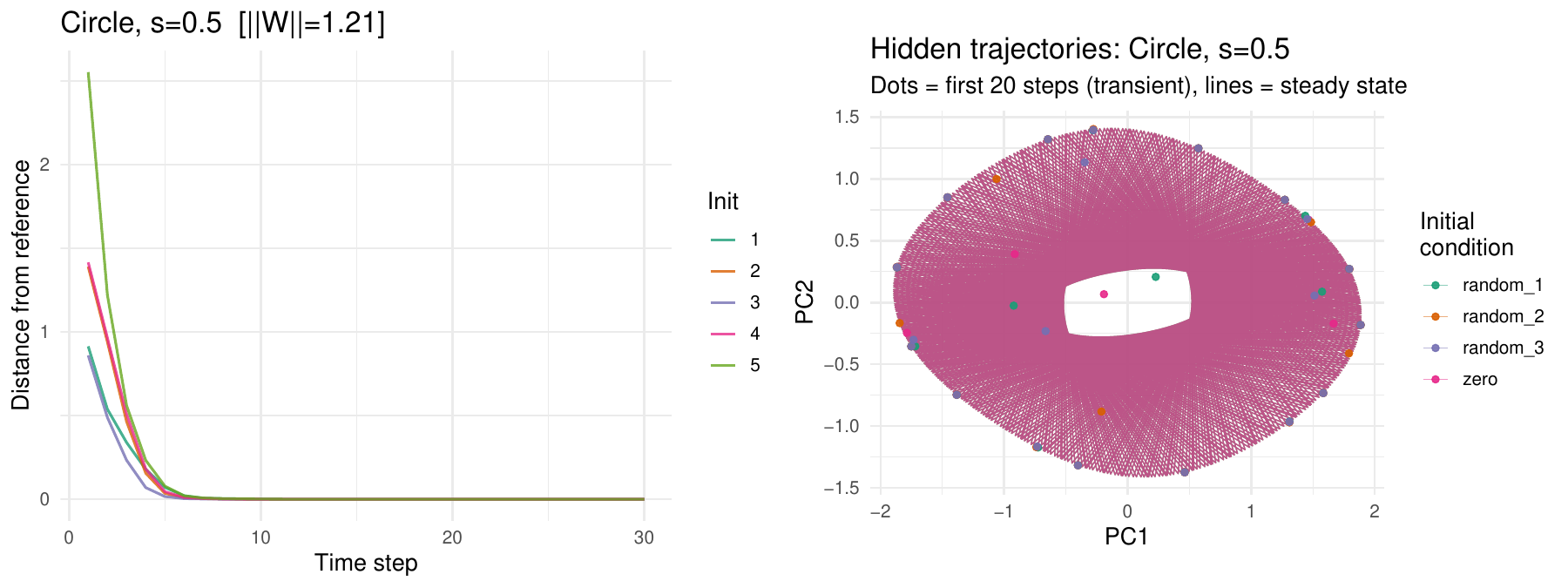}
\includegraphics[width=0.49\textwidth]{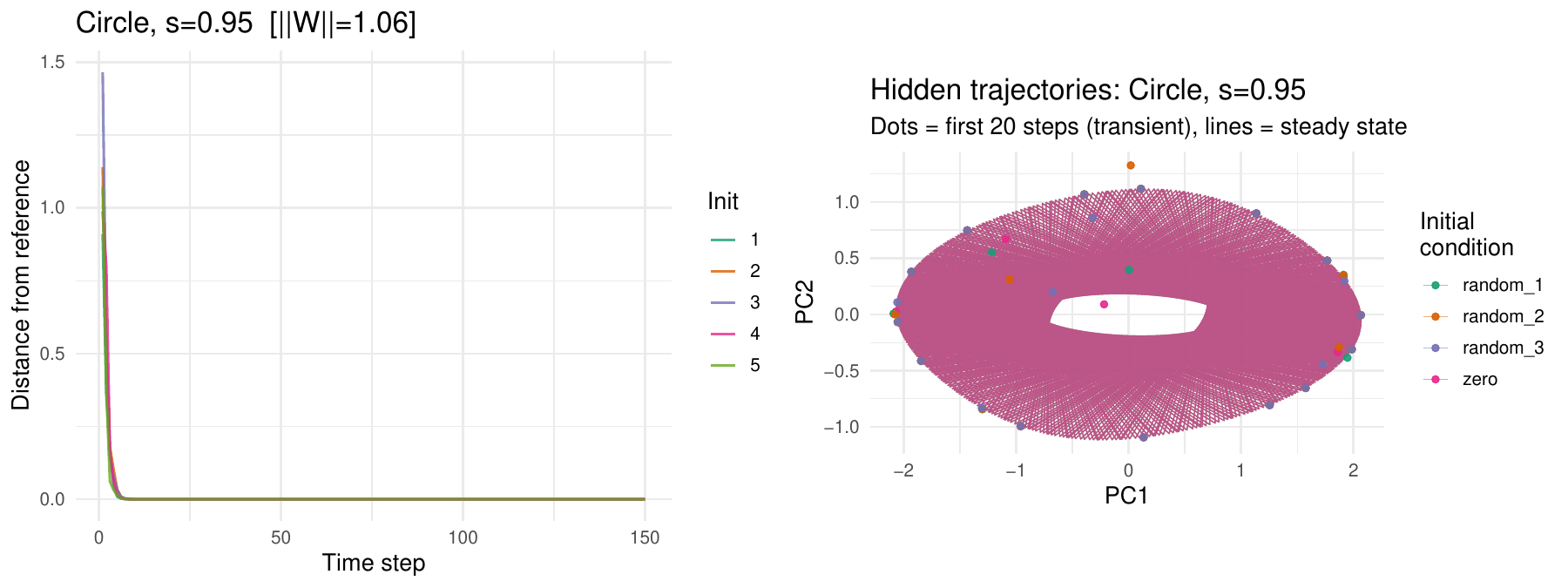}

\medskip

\includegraphics[width=0.49\textwidth]{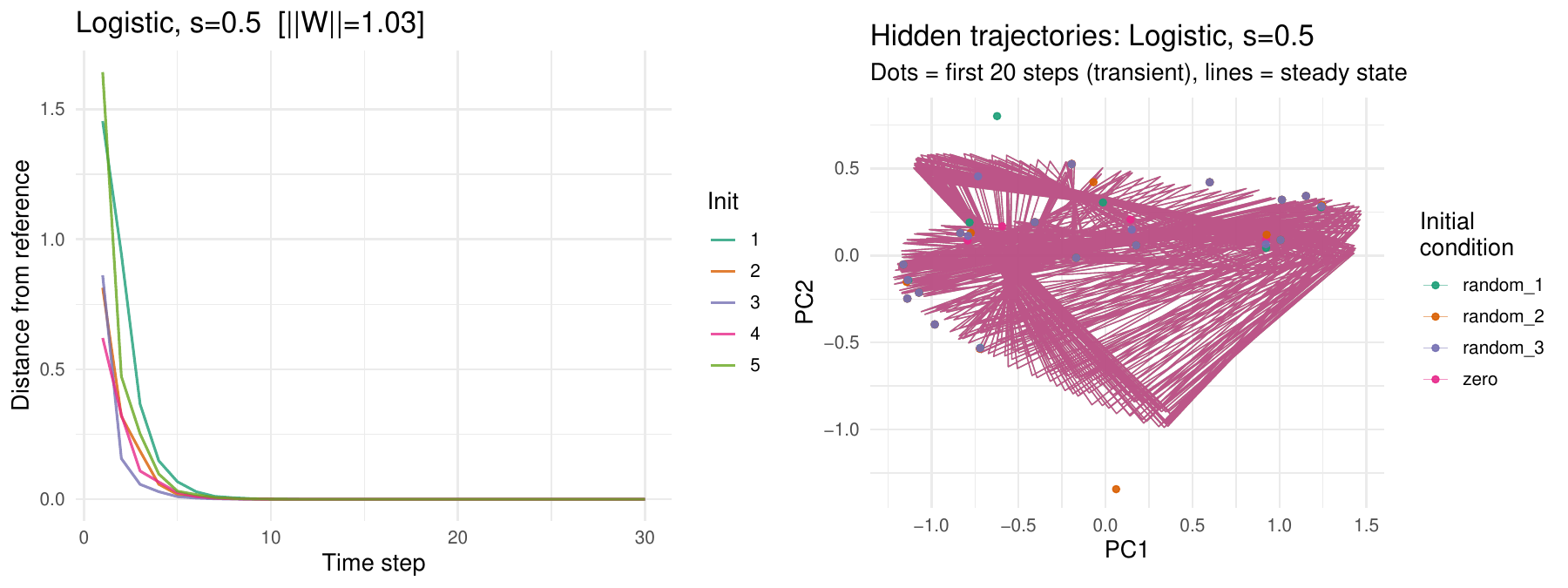}
\includegraphics[width=0.49\textwidth]{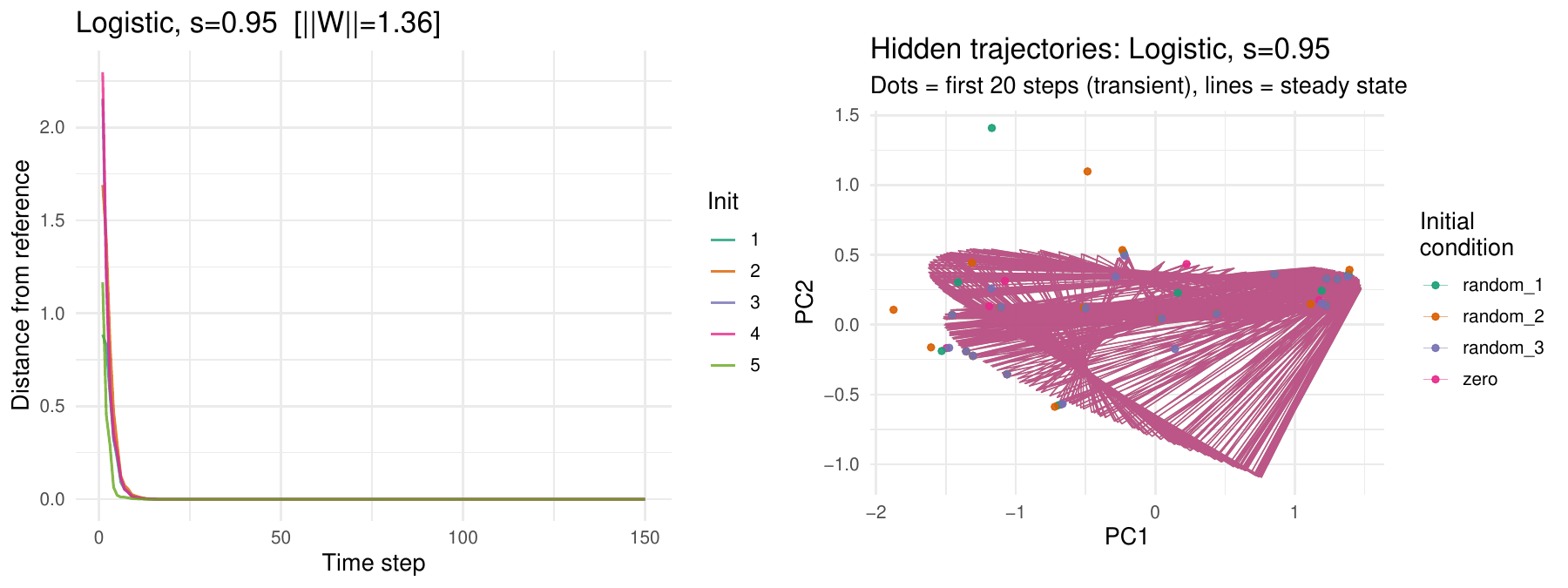}
\caption{Generalized synchronization under four conditions. Each panel shows GS convergence (left: distance from reference trajectory over time for five random initial conditions) and hidden-state geometry (right: PCA projection of trajectories from four initial conditions, with dots marking the transient steps that collapse onto the synchronization set and the line marking the steady state, a continuously traced trajectory on that set (a closed curve for the regular driver, a more intricate attractor for the chaotic one)). Top row: circle driver. Bottom row: logistic driver. Left column: strong initial contraction ($s=0.5$). Right column: weak initial contraction ($s=0.95$). Regular driving with strong contraction yields the fastest convergence. Chaotic driving with weak contraction yields the slowest and noisiest convergence.}
\label{fig:gs_convergence}
\end{figure}

Figure~\ref{fig:gs_convergence} shows the results. Each panel displays two subplots: the left shows the Euclidean distance between the random-initial-condition trajectory and the reference trajectory as a function of time, and the right shows the hidden-state trajectories projected into their first two principal components, with dots marking the first 20 transient steps and lines showing the steady-state regime. The steady state under quasiperiodic driving is itself a moving trajectory: once transients decay, the hidden state rides along the synchronization manifold, a closed curve traced continuously, so it is drawn as a line. The dots mark the initial steps, during which trajectories from different initial hidden states are still off the manifold and collapsing onto it.

Under the regular driver, trajectories converge within about 10 steps when initial contraction is strong and within 50--100 steps when it is weak; under the chaotic driver, convergence is noisier and slower, consistent with the greater dynamical complexity of chaotic driving.

Post-training operator norms exceeded 1 in all four conditions (range: 1.03--1.36), so the sufficient condition for strict ESP no longer holds. Trajectory convergence consistent with generalized synchronization was nevertheless observed in all cases. This is consistent with the theoretical results being sufficient conditions: the phenomena persist beyond the regime in which they are guaranteed.

To test whether the trained networks contract in practice despite $\|W\|_{\mathrm{op}}>1$, we measured the realized contraction along the driven orbit. The fiber Jacobian is $J_t=\operatorname{diag}(1-h_{t+1}^2)W$, and its top conditional Lyapunov exponent $\lambda_{\mathrm{cond}}=\lim_{T\to\infty}\tfrac1T\log\|J_{T-1}\cdots J_0\|$ is the realized mean per-step expansion. Table~\ref{tab:condlyap} reports $\lambda_{\mathrm{cond}}$ for the four trained networks. It is negative in every condition (range $-1.08$ to $-0.78$), including the networks with $\|W\|_{\mathrm{op}}$ as large as $1.36$. Individual per-step Jacobian norms can exceed $1$ (the maximum reaches $1.23$); what contracts is the long product $J_{T-1}\cdots J_0$, not every factor. The $\tanh$ gain $1-h^2\in(0,1]$ shrinks the Jacobian on most steps, so the driven cocycle contracts on average even where the constant bound $\|W\|_{\mathrm{op}}<1$ fails. Trajectories from different initial hidden states converge (residual divergence over the final tenth of the run numerically zero). Driving the same networks with arbitrary white-noise input rather than the trained signal gives the same picture: $\lambda_{\mathrm{cond}}$ stays negative in all four conditions (range $-0.94$ to $-0.72$, column $\lambda_{\mathrm{cond}}^{\mathrm{arb}}$) and trajectories still converge. The operator-norm condition $\|W\|_{\mathrm{op}}<1$ is a conservative worst-case bound, and the realized contraction is not specific to the trained input.

\begin{table}[ht]
\centering
\begin{tabular}{@{}lcccccc@{}}
\toprule
Condition & $\|W\|_{\mathrm{op}}$ & mean $\|J_t\|$ & max $\|J_t\|$ & $\lambda_{\mathrm{cond}}$ & $\lambda_{\mathrm{cond}}^{\mathrm{arb}}$ & GS \\
\midrule
Circle, $s=0.5$    & 1.21 & 0.87 & 1.04 & $-0.78$ & $-0.72$ & yes \\
Circle, $s=0.95$   & 1.06 & 0.80 & 0.86 & $-1.08$ & $-0.94$ & yes \\
Logistic, $s=0.5$  & 1.03 & 0.89 & 0.96 & $-0.83$ & $-0.81$ & yes \\
Logistic, $s=0.95$ & 1.36 & 1.06 & 1.23 & $-0.79$ & $-0.79$ & yes \\
\bottomrule
\end{tabular}
\caption{Realized contraction along the driven orbit for the four trained networks of Experiment~1. The top conditional Lyapunov exponent $\lambda_{\mathrm{cond}}$ is negative in every condition despite post-training operator norms above $1$, and trajectories from different initial conditions converge (GS). Column $\lambda_{\mathrm{cond}}^{\mathrm{arb}}$ reports the same exponent under arbitrary white-noise drive rather than the trained signal; it is also negative in every condition, so the contraction is not specific to the trained input.}
\label{tab:condlyap}

\end{table}

\subsubsection{Experiment 2: Embedding Quality and the Dimension Threshold}

This experiment tests whether the $N > 2d$ threshold from Theorem~\ref{thm:main_synthesis} predicts when distinct driving states remain separated in hidden space. The same quasiperiodic drivers are used (circle, $d=1$; torus, $d=2$) with spectral scale $s = 0.8$ and 100 training epochs. The hidden dimension $N$ varies across conditions: $N \in \{2,3,4,6,8\}$ for the circle and $N \in \{3,4,5,8,12\}$ for the torus. Each condition is run with three random seeds for error bars.

Embedding quality is measured by a collision metric: the fraction of randomly sampled state pairs that are well separated on the base manifold (angular distance $\ge 0.2$ radians for the circle, $\ge 0.4$ for the torus) but nearly coincident in hidden space (Euclidean distance $< 0.1$). Low collision fraction means the representation preserves state separation.

\begin{figure}[ht]
\centering
\includegraphics[width=\textwidth]{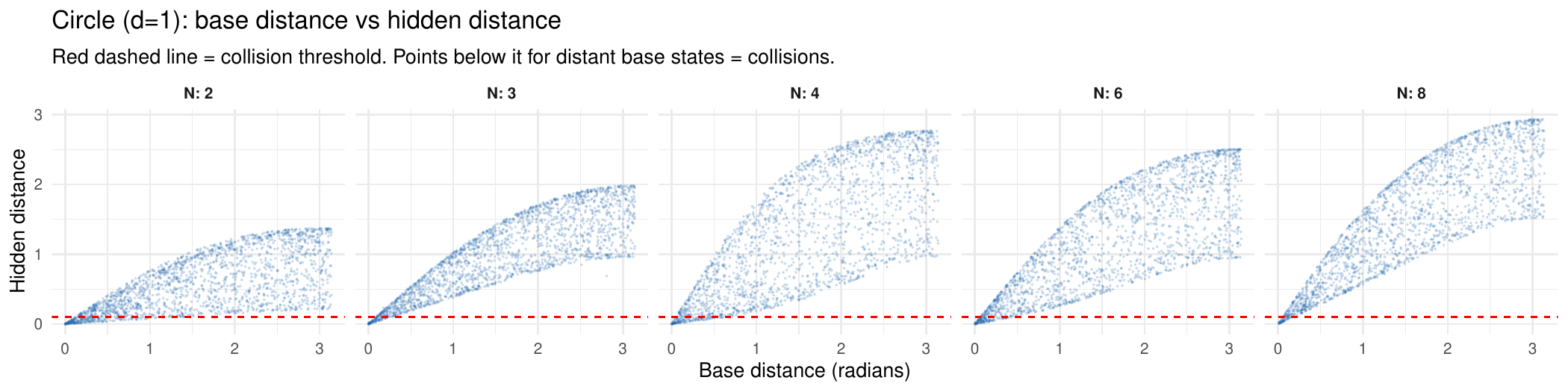}
\caption{Embedding quality for the circle driver ($d=1$) as a function of hidden dimension $N$. Each panel shows 2000 randomly sampled state pairs, with base-manifold distance on the horizontal axis and hidden-space distance on the vertical axis. The red dashed line marks the collision threshold ($\delta_h = 0.1$). Points below this line for distant base states are collisions: distinct states that the representation fails to separate. As $N$ increases, the cloud lifts above the collision threshold.}
\label{fig:embedding_dim}
\end{figure}

\begin{figure}[ht]
\centering
\includegraphics[width=0.6\textwidth]{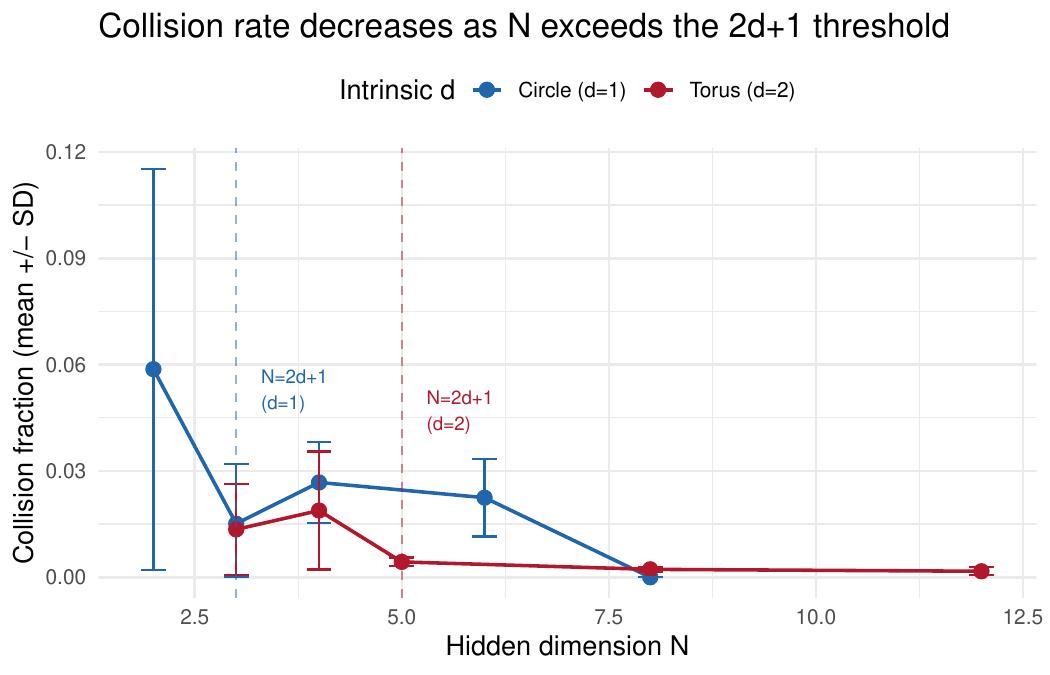}
\caption{Collision rate versus hidden dimension for circle ($d=1$, blue) and torus ($d=2$, red) drivers. Error bars show standard deviation across three random seeds. Vertical dashed lines mark the $N = 2d+1$ threshold from Theorem~\ref{thm:main_synthesis}. Collision rates decrease as $N$ grows, most rapidly near the threshold.}
\label{fig:collision_summary}
\end{figure}

Figure~\ref{fig:embedding_dim} shows the base-manifold distance (horizontal axis) versus hidden-space distance (vertical axis) for each hidden dimension. As $N$ increases past the $2d+1$ threshold, the point cloud lifts away from the horizontal axis: states that are far apart on the manifold become far apart in hidden space. Figure~\ref{fig:collision_summary} summarizes the collision rates across conditions. Collision fractions decrease as $N$ increases, with the steepest reduction occurring near the theoretical thresholds $N = 3$ (for $d=1$) and $N = 5$ (for $d=2$).

\begin{figure}[ht]
\centering
\includegraphics[width=\textwidth]{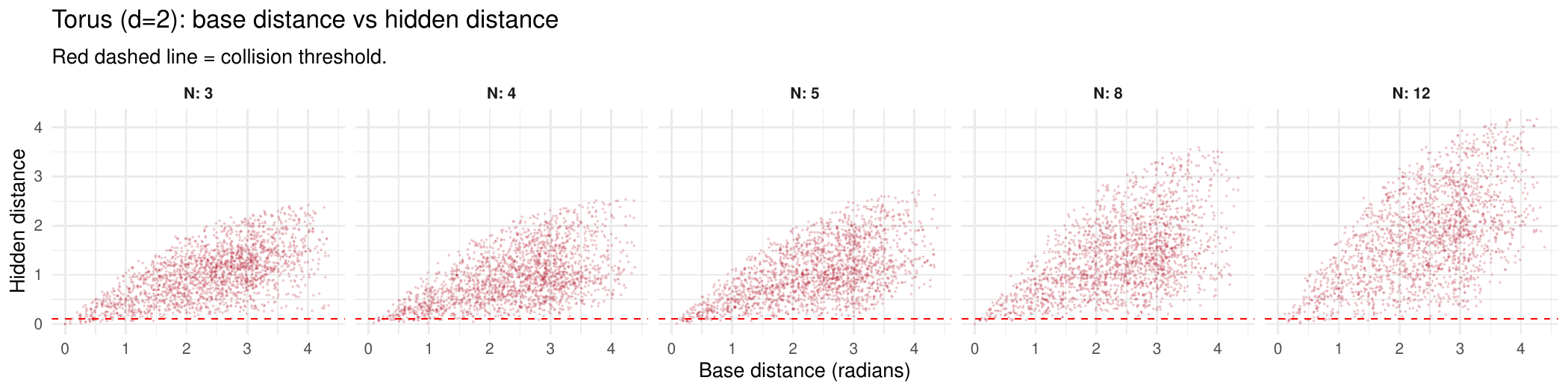}
\caption{Embedding quality for the torus driver ($d=2$) as a function of hidden dimension $N$. Same format as Figure~\ref{fig:embedding_dim}. The torus requires larger $N$ to avoid collisions ($N \ge 5$ versus $N \ge 3$ for the circle), consistent with the $N > 2d$ scaling.}
\label{fig:embedding_dim_torus}
\end{figure}

\subsubsection{Experiment 3: Prediction Accuracy Constrains Representational Geometry}

This experiment illustrates the prediction-separation link (Proposition~\ref{prop:prediction_separation}). The proposition's hypotheses (uniform prediction bound, known separation modulus) are not directly verified; the experiment tests whether the qualitative relationship between prediction accuracy and state separation holds in trained networks. Three separate $\tanh$ RNNs ($N=8$) are trained on the same circle driver with different initial contraction strengths: $s = 0.3$ (strong), $s = 0.7$ (moderate), and $s = 0.95$ (weak). All three are trained for 150 epochs. A ridge-regression readout is trained on the first half of the trajectory to predict $K=5$ future observations from the hidden state, and evaluated on the second half.

Mean prediction errors on the held-out data are 0.065 (strong), 0.036 (moderate), and 0.029 (weak). The supremum-to-mean error ratio, computed from the unrounded errors, is $2.70$, $2.49$, and $1.99$ respectively (sample suprema $0.175$, $0.090$, $0.057$), so the observed test-set supremum runs about two to three times the mean. This supremum is a finite test-set maximum for the $K$-step readout, not the supremum over $M$ that Proposition~\ref{prop:prediction_separation} assumes. Training controls expected rather than uniform error, so the probabilistic relaxation in Corollary~\ref{cor:probabilistic_separation} is the appropriate guarantee; we do not estimate its $G$ and $p$ here. The less contractive networks achieve lower prediction error, possibly reflecting greater representational capacity. Post-training operator norms are 0.72 (strong, ESP holds), 1.27 (moderate, ESP not guaranteed), and 1.48 (weak, ESP not guaranteed). The theory's sufficient conditions (strict ESP) hold only for the strong case, but the prediction-separation relationship is visible across all three.

\begin{figure}[ht]
\centering
\includegraphics[width=0.65\textwidth]{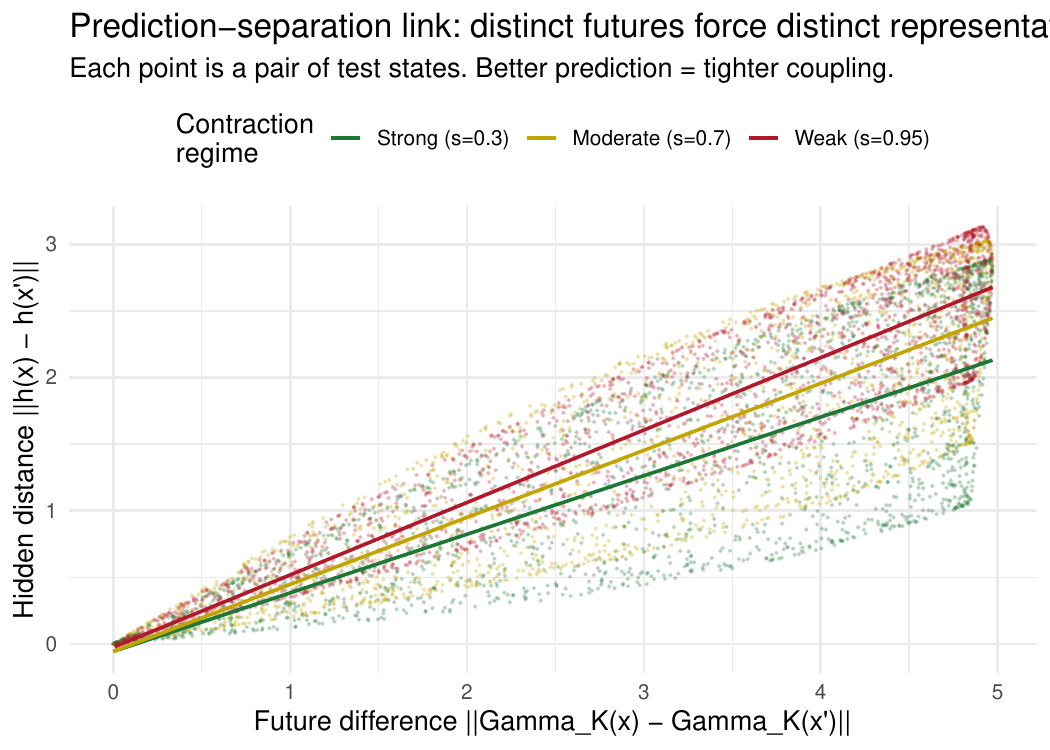}
\caption{Prediction-separation link. Each point represents a pair of test states. Horizontal axis: difference in their $K=5$-step futures. Vertical axis: difference in their hidden representations. Colors indicate three contraction regimes. Regression lines show that networks with lower prediction error (less contractive, more expressive) produce tighter coupling between future differences and representational differences. The relationship holds even when the sufficient condition for strict ESP no longer holds after training.}
\label{fig:pred_sep}
\end{figure}

Figure~\ref{fig:pred_sep} shows the core result. For randomly sampled pairs of test states, the horizontal axis measures how different their futures are ($\|\Gamma_K(x) - \Gamma_K(x')\|$) and the vertical axis measures how different their hidden representations are ($\|h(x) - h(x')\|$). States with distinct futures tend to have distinct representations. The regression lines show that the coupling is tightest for the network with the lowest prediction error.

\begin{figure}[ht]
\centering
\includegraphics[width=0.6\textwidth]{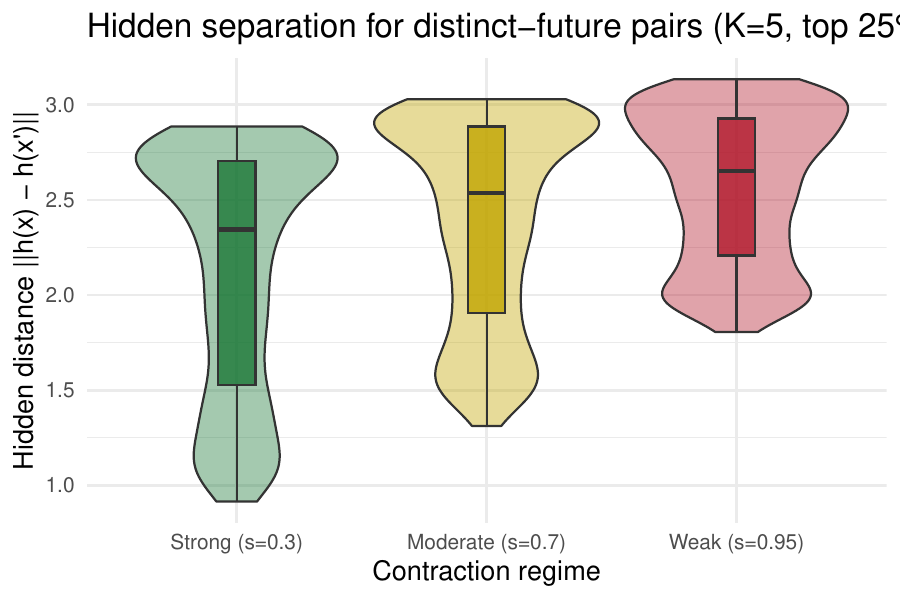}
\caption{Distribution of hidden-state separation for pairs of test states whose futures differ substantially (top quartile of $\|\Gamma_K(x) - \Gamma_K(x')\|$). Violin width shows the density of hidden distances $\|h(x) - h(x')\|$ within each regime. The strongly contractive network (s=0.3) has lower median separation than the weakly contractive networks, consistent with a smaller hidden-state dynamic range. All three regimes separate distinct-future states, but the distribution shifts upward with representational capacity.}
\label{fig:pred_sep_violin}
\end{figure}

\subsubsection{Experiment 4: Manifold Topology Recovery}

This experiment tests whether trained RNNs develop hidden manifolds with topology matching that of the driving signal. An $N=8$ network is trained on a head-direction-like signal on $S^1$ (150 epochs), and an $N=12$ network is trained on a two-frequency signal on $T^2$ (200 epochs).

\begin{figure}[ht]
\centering
\includegraphics[width=0.55\textwidth]{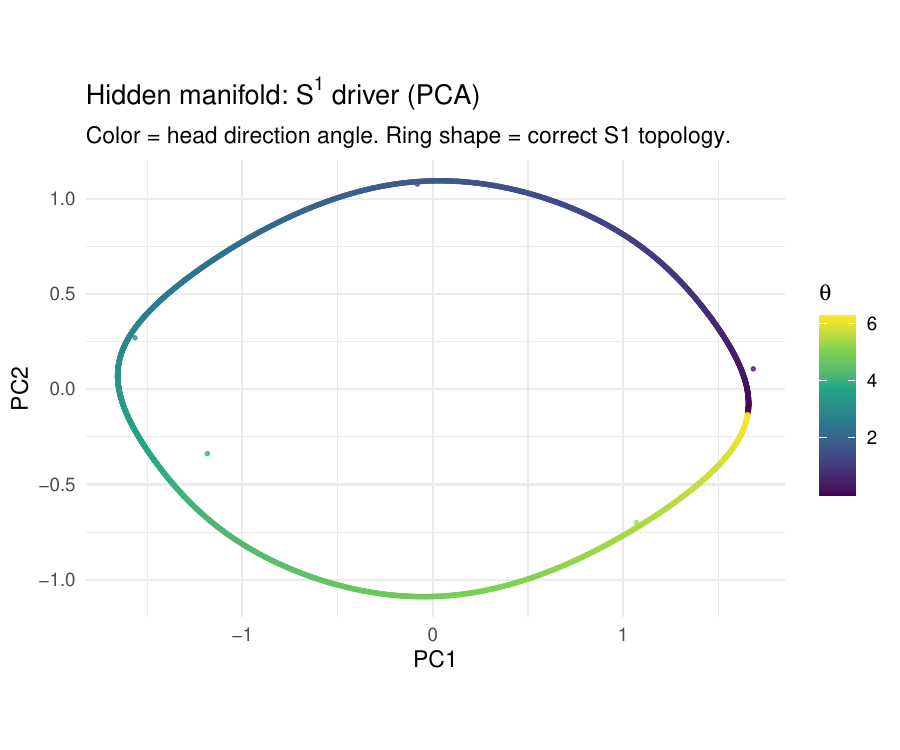}
\caption{Hidden manifold for the $S^1$ head-direction driver. First two principal components of the hidden state, colored by the driving angle $\theta$. The ring shape and smooth color gradient are consistent with preservation of the topology and metric structure of $S^1$. Three PCs capture 99.8\% of the variance.}
\label{fig:manifold_ring}
\end{figure}

Figure~\ref{fig:manifold_ring} shows the first two principal components of the hidden states for the $S^1$ driver, colored by the driving angle $\theta$. The hidden manifold forms a ring, and the color gradient wraps smoothly around it, consistent with preservation of both the topology (a circle) and the metric structure (nearby angles map to nearby hidden states) of the driving dynamics. The first three PCs capture 99.8\% of the variance (76.6\%, 22.9\%, 0.4\%), consistent with a one-dimensional manifold embedded in two effective dimensions.

\begin{figure}[ht]
\centering
\includegraphics[width=\textwidth]{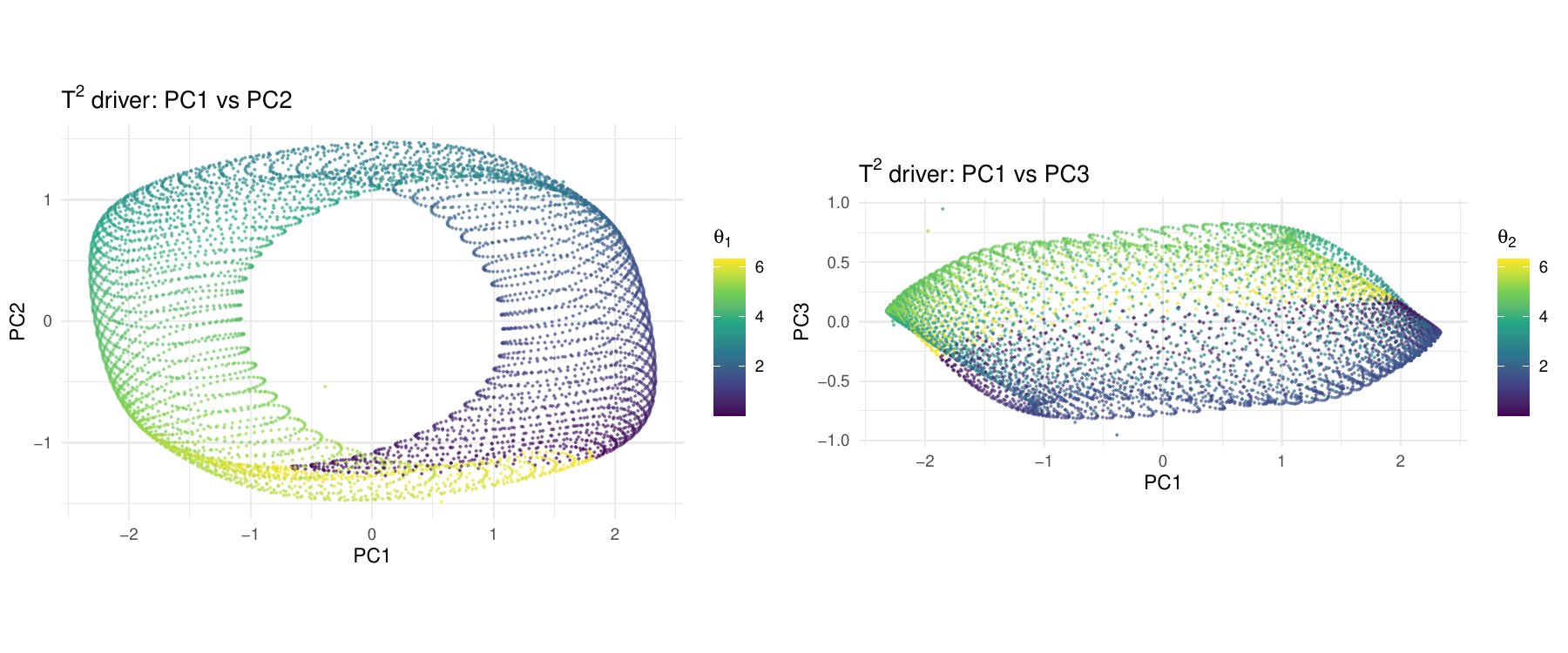}
\caption{Hidden manifold for the $T^2$ two-frequency driver. Left: PC1 versus PC2, colored by $\theta_1$. Right: PC1 versus PC3, colored by $\theta_2$. Each frequency is resolved in a different principal component pair, consistent with a two-dimensional embedding. Four PCs capture 99.6\% of the variance.}
\label{fig:manifold_torus}
\end{figure}

Figure~\ref{fig:manifold_torus} shows two PCA projections of the hidden states for the $T^2$ driver: PC1 versus PC2 (colored by $\theta_1$) and PC1 versus PC3 (colored by $\theta_2$). Each frequency is resolved in a different principal component pair, consistent with a two-dimensional manifold. The first four PCs capture 99.6\% of the variance.

To assess whether the manifold topology is detectable from the hidden states, we applied persistent homology, a method that identifies topological features (loops, voids) in a point cloud at multiple spatial scales, in two protocols. In the point-wise protocol, 300 hidden states are subsampled randomly (ignoring temporal order) and Vietoris-Rips persistent homology is computed on the resulting point cloud. In the sequential protocol, 300 consecutive hidden states are taken in trajectory order. For the $S^1$ driver, both protocols detect a single dominant $H_1$ bar (persistence gap $= \infty$), indicating a ring. For the $T^2$ driver we apply a surrogate-null criterion: a bar is genuine if its persistence exceeds the 95th percentile of a matched null (Gaussian point clouds matched in sample size, dimension, and per-coordinate mean and standard deviation for the point-wise protocol; per-coordinate Fourier phase-randomized series for the sequential protocol). Point-wise homology gives two $H_1$ bars above the null (longest-to-second-longest gap $3.1$), matching the two independent cycles of a torus, but no $H_2$ bar above the null (gap $\approx 1.1$): the toroidal void is not separated from noise at $300$ points in a four-dimensional projection. The sequential protocol gives one $H_1$ bar above the null and no $H_2$. The raw counts (67 $H_1$ and 8 $H_2$ bars point-wise) are dominated by short-lived noise; the genuine features are those exceeding the null.

\begin{figure}[ht]
\centering
\includegraphics[width=0.85\textwidth]{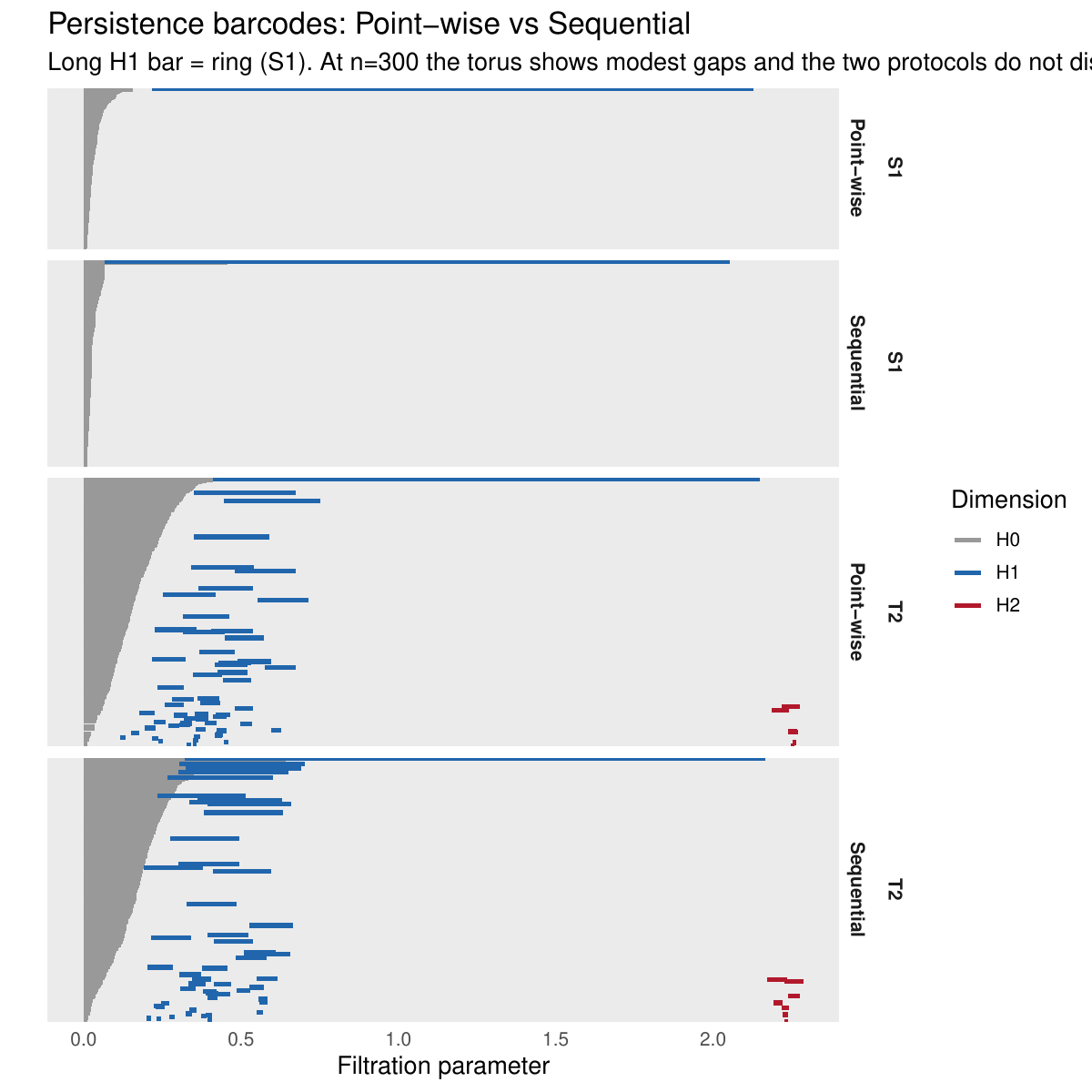}
\caption{Persistence barcodes for point-wise and sequential protocols on both drivers. Each horizontal bar represents a topological feature: its left endpoint is the filtration scale at which it appears (birth) and its right endpoint is the scale at which it disappears (death). Long bars indicate genuine topological structure; short bars indicate noise. For $S^1$, both protocols detect a single long $H_1$ bar (a loop), consistent with ring topology. For $T^2$, two point-wise $H_1$ bars exceed the 95th-percentile surrogate-null threshold (the two cycles of a torus), while no $H_2$ bar does; the sequential protocol gives one $H_1$ bar above the null. At 300 points the two protocols do not dissociate.}
\label{fig:ph_barcodes}
\end{figure}

\section{Discussion}
\label{sec:discussion}

\subsection{Connection to Neural Coding and Population Dynamics}
\label{subsec:neural_coding}

Section~\ref{subsec:sensory_problem} introduced the neural manifold hypothesis as one of three research programs motivating this work. Here we develop the connection in detail. The question is: how do the low-dimensional manifolds observed in sensory cortex \citep{CunninghamYu2014, Gallego2017, Gallego2018, Gallego2020, Kim2017, Chaudhuri2019, Gardner2022} arise? The present results provide a formal mechanism. The sensory signals that instantiate the regular regime are catalogued in Section~\ref{subsec:sensory_regular}.

When a recurrent circuit satisfies the contraction conditions developed in Section~\ref{sec:main_results} and receives low-dimensional regular input, the synchronization map $f$ embeds the driving manifold smoothly into neural state space, provided the hidden dimension exceeds $2d$ and the network and observation are generic (Theorem~\ref{thm:main_synthesis}). The image $f(M)$ is a neural manifold in the empirical sense, but here its existence and geometry follow from explicit sufficient conditions rather than from post hoc dimensionality reduction. The conditions are not necessary; the numerical experiments show that manifold recovery persists beyond the strict contraction regime. The conditions instead identify one setting in which the representational structure can be characterized sharply. The dimension bound $N > 2d$ connects directly to the distinction between intrinsic and embedding dimensionality drawn by \citet{JazayeriOstojic2021}: $d$ is the intrinsic dimension of the sensory dynamics; $N$ is the embedding dimension in neural state space.

\citet{Kim2017} and \citet{Chaudhuri2019} confirmed ring-shaped attractor manifolds for head direction in \emph{Drosophila} and mice, respectively. Theorem~\ref{thm:main_synthesis} provides sufficient conditions for a driven recurrent circuit to produce exactly such a ring when receiving head-direction signals on $S^1$. Experiment~4 (Section~\ref{subsec:numerical_experiments}) shows that trained networks develop ring- and torus-like hidden manifolds consistent with these predictions.

\citet{VyasEtAl2020} framed neural computation as the temporal evolution of population state on manifolds. The present framework shows that the manifold itself arises from the synchronization between recurrent dynamics and the environment. The ``computation through dynamics'' picture and the embedding picture are complementary: the former describes what happens on the manifold, while the latter describes why the manifold exists and what it preserves.

\subsection{Connection to Predictive Processing}
\label{subsec:predictive_processing}

Section~\ref{subsec:sensory_problem} identified predictive processing as the second research program motivating this work. The question deferred there is how the embedding results give that framework a geometric grounding. In predictive processing, cortical hierarchies operate by having each level generate predictions about the activity of the level below. Only the discrepancy between prediction and observation propagates forward \citep{RaoBallard1999, Friston2010, clark2013}. Backward connections carry predictions. Forward connections carry prediction errors. The gain on these error signals is modulated by their estimated precision, a measure of signal reliability encoded as the inverse variance of the error distribution \citep{Feldman2010}.

The present results complement this account in two ways.

First, the synchronization relation $f(\phi(x)) = F(f(x), \omega(x))$ identifies the predicted next neural state with the actual next neural state after processing the current observation. The embedding is prediction, realized as dynamical structure. If the recurrent dynamics of predictive processing circuits satisfy contraction conditions, with the dimension and genericity conditions of Theorem~\ref{thm:main_synthesis}, then the internal model that minimizes prediction error would be a smooth embedding of the sensory dynamics. Whether this holds for specific predictive processing architectures is an open question (Section~\ref{subsec:limitations}).

Second, the prediction-separation result (Proposition~\ref{prop:prediction_separation}) formalizes a core intuition of predictive processing. States whose futures differ should be represented differently; states whose futures are identical need not be distinguished. The proposition makes this quantitative: prediction error $\varepsilon$ sets a resolution limit below which states can collapse, and the separation modulus $\eta$ of the forward observation map determines which distinctions are preserved at each scale.

Friston's generalized filtering framework implements Bayesian inference as a set of ordinary differential equations. The state vector in that framework comprises not just the hidden state but also its velocity, acceleration, and higher temporal derivatives \citep{Friston2010GF, DaCostaEtAl2025}. Whether these filtering dynamics satisfy the contraction conditions required by Theorem~\ref{thm:main_synthesis} is an open question. An affirmative answer would formally connect the Bayesian and geometric accounts of sensory representation. The precision-weighting mechanism central to attention in predictive processing \citep{Feldman2010} maps naturally onto gain modulation of prediction error signals; whether this mechanism provides a biological substrate for the contraction required by the echo state property is a specific target for future analysis.

\subsection{Relation to Empirical Findings}
\label{subsec:empirical}

Section~\ref{subsec:sensory_problem} introduced the \citet{UribarriMindlin2022} and \citet{Ostrow2024} results as evidence that trained recurrent networks develop topologically faithful internal representations. Here we ask what those findings, obtained under chaotic driving, imply for the regular regime. A strange attractor has fractal geometry and sensitive dependence on initial conditions. In that setting, bunching and fractal-dimension issues are in play. The regular regime avoids these obstructions and admits cleaner guarantees. The conditions developed here are easier to satisfy for regular dynamics than for chaotic dynamics, so the empirical success of embedding under chaotic driving is consistent with the expectation that embedding should hold a fortiori for the regular case.

Many chaotic attractors contain a dense set of unstable periodic orbits: trajectories that return to their starting point but diverge from it under small perturbation. These orbits carry the attractor's topological structure \citep{GilmoreLetellier2007}. The \citet{UribarriMindlin2022} finding that trained LSTMs preserved the topological organization of periodic orbits suggests that embedding a strange attractor in practice amounts to embedding its periodic orbit skeleton, which is regular structure. This connection is interpretive, not proved here, but it explains why the regular-dynamics results may be relevant beyond the strictly quasiperiodic regime. 
Beyond the LSTM results of \citet{UribarriMindlin2022}, \citet{Ostrow2024} demonstrated the same phenomenon in transformers and state-space models trained on partially observed dynamical systems: architectures with stronger inductive bias for temporal recurrence produced better embeddings. In a biological setting, \citet{Raut2025} showed that delay embedding of a single scalar observable (pupil diameter) reconstructs the spatiotemporal dynamics of neuronal calcium, metabolism, and hemodynamics across the mouse cortex, providing direct in vivo evidence that temporal embedding of low-dimensional signals recovers high-dimensional brain dynamics.

\citet{Sono2026} provide the most direct biological validation of the reservoir computing framework used here. They cultured rat cortical neurons in microfluidic devices that imposed modular connectivity, integrated the cultures with high-density microelectrode arrays, and trained the system in real time via FORCE learning with closed-loop feedback. The cultured networks learned to generate periodic temporal signals (sine, triangle, and square waves) and approximated chaotic (Lorenz) dynamics. Two findings are relevant. First, modular connectivity was required: homogeneous cultures exhibited excessive synchronization and failed to learn, while lattice and hierarchical architectures succeeded. One interpretation within the present framework is that excessive synchronization collapses the effective hidden dimension (all neurons lock together, reducing the independent degrees of freedom), while modular connectivity distributes activity across independent channels, increasing the effective $N$. Second, PCA trajectory analysis showed that closed-loop feedback transformed irregular high-dimensional spontaneous activity into structured low-dimensional trajectories, consistent with the generalized synchronization regime in which the driven network's state becomes a deterministic function of the driving signal. Regular target signals (periodic waves) were learned and sustained after training was halted; the chaotic target degraded post-training, consistent with the prediction that regular dynamics are the regime in which faithful embedding is most robust.

\subsection{What the Experiments Reveal Beyond the Theory}
\label{subsec:theory_vs_experiment}

The theorems in this paper provide sufficient conditions for faithful embedding: strict ESP, the bunching condition, $N > 2d$, and genericity of the observation function and network. The numerical experiments (Section~\ref{subsec:numerical_experiments}) test these conditions and reveal where the sufficient conditions are conservative.

The most consistent finding is that training by gradient descent pushes the operator norm of the recurrent weight matrix above 1. When this happens, the sufficient condition for strict ESP no longer holds, and the theorems no longer guarantee embedding. Post-training operator norms ranged from 0.72 to 1.48 across experiments; the networks whose operator norm exceeded 1 fell in the range 1.01--1.48, while the most strongly contractive networks stayed below 1. The sufficient condition held after training only for the most strongly contractive initialization ($s = 0.3$ in Experiment~3, and most networks in Experiment~2). Generalized synchronization, manifold recovery, and the prediction-separation relationship were nevertheless observed in every condition tested, including those where the sufficient condition failed.

This gap between theory and experiment is informative. The theorems characterize a regime in which embedding is guaranteed. The experiments show that the phenomena extend well beyond that regime. The sufficient conditions are not necessary conditions. Gradient descent on a one-step prediction objective finds network configurations that support generalized synchronization even when the operator norm exceeds 1. The conditional-Lyapunov analysis of Experiment~1 (Table~\ref{tab:condlyap}) shows why: the realized contraction along the driven orbit is negative even when $\|W\|_{\mathrm{op}}>1$, because the $\tanh$ gain $1-h^2\in(0,1]$ shrinks the Jacobian on most steps. This contraction holds under arbitrary white-noise drive as well as under the trained signal (Table~\ref{tab:condlyap}). The operator-norm condition is a conservative worst-case bound; the realized contraction is not specific to the trained input.

Experiment~3 produced a counterintuitive result. Networks initialized with weaker contraction ($s = 0.95$) achieved lower prediction error (mean $\varepsilon = 0.029$) than networks initialized with stronger contraction ($s = 0.3$, mean $\varepsilon = 0.065$). One possible explanation is that weaker contraction allows the hidden state to maintain richer dynamics instead of collapsing rapidly toward a fixed function of the input. To distinguish this from a capacity ceiling at $N=8$, we computed the participation ratio (effective rank) of the hidden-state covariance, $\mathrm{PR}=(\sum_i\lambda_i)^2/\sum_i\lambda_i^2$. It rises as contraction weakens: $1.19$ (strong), $1.34$ (moderate), $1.62$ (weak), all well below the ceiling $N=8$. The weakly contractive network uses more of its state space, which is consistent with the richer-dynamics account and not with a capacity limit. We present this as a diagnostic, not a decisive test. The theory guarantees embedding under strong contraction. The experiments indicate that the representations formed under weak contraction, while not covered by the present theorems, produce at least as good prediction-separation coupling and often better.

Experiment~4 found that trained networks develop hidden manifolds consistent with the expected topology: a clear ring for $S^1$, and partial torus structure for $T^2$. Persistent homology detected a single dominant $H_1$ bar for the circle driver in both point-wise and sequential protocols. For the torus, two $H_1$ bars exceed the surrogate null (the two cycles of a torus), but the $H_2$ void does not, reflecting the difficulty of recovering the toroidal void from low-dimensional PCA projections of 300 points (Section~\ref{subsec:numerical_experiments}). Larger samples and higher-dimensional projections may recover it.

Five further analyses, reported in Appendix~\ref{app:experiments}, test the robustness of these findings and isolate what training contributes. The embedding persists at large hidden dimension (Experiment~5), occurs in frozen reservoirs with untrained recurrent weights (Experiment~6), generalizes to novel trajectories, and survives recovery after a chaotic excursion displaces the hidden state (Experiment~7). Training reshapes rather than creates the synchronization regime: it redistributes contraction across more hidden directions without speeding synchronization (Experiment~8) and shapes the representational geometry for the training objective while leaving the embedding intact (Experiment~9).

\subsection{Implications for Sensory Representation}

The results place a mathematically explicit account of sensory representation within reach of known cortical population dynamics. Evidence from sensory cortex suggests input-dependent contraction: fading memory on timescales of 100--300\,ms, variability quenching at stimulus onset, and smooth population responses to nearby stimuli (Section~\ref{subsec:driven_systems}).

These observations give an order-of-magnitude estimate of the contraction parameters for one circuit. For head-direction cells, a fading-memory horizon of about 100--300\,ms at a $\sim$10\,ms step corresponds to a contraction rate $\rho \approx e^{-1/\text{horizon}} \approx 0.90$--$0.97$. The conjugacy condition number $\kappa$ measures the departure of the angular drive from a rigid rotation; for near-uniform tuning we estimate $\kappa \lesssim 1.1$ as a rough proxy. With $1/\kappa \approx 0.91$, the representative value $\rho \approx 0.95$ violates the strict bunching condition $\rho < 1/\kappa$, and the plausible range straddles the boundary, placing such circuits at the edge of the strictly guaranteed regime. This is consistent with the experiments, in which embedding persists past the sufficient conditions. The contraction parameter is not a reservoir-computing artifact: contraction is certifiable in biologically motivated recurrent models, including Hopfield and firing-rate networks with symmetric weights \citep{Centorrino2023} and with Hebbian and anti-Hebbian dynamic synapses expressed in circuit-level quantities such as neural and synaptic decay rates and maximum in-degree \citep{Centorrino2024, Kozachkov2020}. These certify biologically motivated models, not biological tissue, and not any specific sensory circuit. The theorems identify sufficient conditions under which a recurrent circuit builds a smooth internal model of the low-dimensional dynamics that drive it. The experiments show that gradient-descent-trained $\tanh$ RNNs exhibit the predicted embedding behavior even when the sufficient condition for strict ESP no longer holds after training.

The representational burden scales with the intrinsic dimension of the sensed dynamics. A sensory circuit needs enough degrees of freedom to preserve the structure of the signal it tracks, and that number is set by the signal's dimensionality, not by the complexity of the full environment. This places the relevant level of analysis at the mesoscale of recurrent population dynamics rather than at individual neurons.

The synchronization map $f: M \to \R^N$ gives the learned correspondence between states of the sensory stream and states of the recurrent circuit. Low prediction error forces a scale-limited embedding, not a perfect one. The resolution limit and its perceptual consequences are characterized by Proposition~\ref{prop:prediction_separation}.

A theory of sensory representation based on state-space embeddings has been proposed \citep{OReillyShah2025, OReillyShahNDPLS}; the present results supply mathematical support for its contractive-regime component.

\subsection{Limitations and Open Problems}
\label{subsec:limitations}

The theorems assume fixed network parameters and strict uniform ESP. The experiments use gradient-descent training on standard $\tanh$ RNNs, for which a simple sufficient condition for ESP is $\|W\|_{\mathrm{op}} < 1$ at initialization \citep{Jaeger2001, LukoseviciusJaeger2009}. Training changes $W$, and in most of our experiments the post-training operator norm exceeded 1. The theory does not currently account for this: it characterizes the fixed-weight regime, not the training trajectory. Extending the results to cover networks whose weights evolve during training, or proving that gradient descent on a prediction objective converges to a regime where input-dependent ESP holds, is the most pressing open problem raised by the experiments.

Noise robustness is not addressed \citep{StarkBroomhead2003}. The regular regime is a mathematically tractable baseline; extending the same bounds to chaotic or stochastic inputs is future work. Input-dependent or local ESP variants \citep{ManjunathJaeger2013, YildizJaegerKiebel2012} are more realistic for biological sensory circuits and would bring the theory closer to what the experiments demonstrate. Gated architectures such as long short-term memory and gated recurrent unit networks admit related sufficient stability conditions, but these depend on architecture-specific inequalities involving the weights and gates \citep{MillerHardt2018, BonassiFarinaScattolini2021}; extending the present embedding results to those architectures is a separate line of work.

Systems with multiple coexisting stable states (multistability), attractor competition, and persistent state-dependent activity all point toward settings in which strict ESP fails. Those cases fall outside the scope of the present theorems. The value of the current analysis is that it identifies one regime in which representational structure can be characterized sharply, and it makes clear what must be relaxed when moving beyond that regime.

Four specific open problems deserve mention. First, whether the generalized filtering dynamics of the free energy framework \citep{Friston2010GF, DaCostaEtAl2025} satisfy contraction conditions compatible with Theorem~\ref{thm:main_synthesis}. Second, whether the precision-weighting mechanism of predictive processing provides a biological implementation of the contraction needed for the echo state property. Third, extending the collision-metric analysis of the numerical experiments (Section~\ref{subsec:numerical_experiments}) to real neural population recordings from sensory cortex, testing whether state separation scales with hidden dimensionality and prediction accuracy as the theory predicts. Fourth, characterizing the class of training objectives and architectures for which gradient descent reliably produces networks in the embedding regime, even when strict ESP is not maintained.

\subsection{Testable Predictions}

The theory and experiments together generate predictions for neural population recordings and computational models. We separate predictions already confirmed in silico by our experiments from novel predictions for real neural data. Confirmed here: the prediction-separation coupling (Experiment~3), the graceful degradation of separation with $N$ (Experiment~2), and cleaner geometry under regular than chaotic driving (Experiment~1). The novel predictions are that these same relationships hold in neural recordings, together with those below. First, prediction accuracy should correlate with representational separation: populations that predict sensory dynamics more accurately should exhibit better state separation in their hidden manifolds, regardless of whether strict contraction conditions hold. Experiment~3 is consistent with this for trained RNNs; the prediction is that the same relationship holds in neural data. Second, state separation should degrade gradually as hidden dimensionality decreases below $2d+1$, not catastrophically. For delay-coordinate maps, the self-intersection theorem of \citet{SauerYorkeCasdagli1991} shows that the failure set has dimension at most $2d - N$. If this pattern transfers to synchronization maps, failures should be sparse and localized rather than global. Experiment~2 confirms this pattern: collision fractions increase smoothly as $N$ decreases rather than jumping at a threshold. Third, regular or quasiperiodic drives should produce cleaner representational geometry than chaotic drives (Experiment~1). Fourth, modular network connectivity should support embedding where homogeneous connectivity fails, as demonstrated biologically by \citet{Sono2026}.

The theory also predicts specific forms of representational distortion. Frequently encountered distinctions should be represented more finely than rare ones. States with near-identical futures should be the first to collapse together. Resolution limits should appear as graded failures of state separation rather than as wholesale loss of structure.

\section{Conclusion}
\label{sec:conclusion}
We began from a concrete question: why do sensory cortical populations organize their activity on low-dimensional manifolds that track behaviorally relevant variables, and how does prediction shape the geometry of those manifolds? Our analysis provides a mechanistic answer in one tractable regime. When a recurrent circuit is (input-)contracting and is driven by low-dimensional regular sensory dynamics, such as rotations on a circle or quasiperiodic motion on a torus, its long-run state becomes a smooth function of the sensory state. Under mild conditions, this synchronization map is a $C^{1}$ embedding: the circuit constructs an internal manifold $f(M)\subset\mathbb{R}^{N}$ that preserves the topology and local geometry of the sensory manifold $M$. The representational burden scales with the intrinsic dimension $d$ of the effective sensory dynamics, not with the complexity of the external world: a hidden dimension $N>2d$, compatible with Takens' Theorem, already suffices for faithful embedding in this regime when the network and observation are generic.

Beyond existence, we showed that prediction accuracy constrains representational geometry even when contraction properties are unknown. A circuit that predicts future sensory inputs with small error cannot arbitrarily collapse states with different futures; instead, it realizes a scale-limited embedding whose resolution is set by the prediction error and by how strongly futures separate on the sensory manifold. This has direct perceptual consequences: categorical boundaries, metameric equivalence of physically distinct stimuli, and discrimination thresholds all emerge as natural failure modes of an otherwise faithful predictive embedding.

Specializing generalized synchronization and modern embedding theorems to regular dynamics on circles and tori made these statements explicit and checkable. Our numerical experiments with trained $\tanh$ recurrent networks support the theoretical picture: networks driven by head-direction-like and multi-frequency inputs develop ring- and torus-like hidden manifolds; state separation improves most rapidly near the $2d+1$ threshold; and accurate predictors exhibit strong coupling between differences in future inputs and differences in hidden state. At the same time, training routinely drives recurrent weights beyond the strict contraction regime required by our theorems, yet convergence consistent with generalized synchronization and manifold recovery persist. This gap highlights that our conditions are sufficient but not necessary, and points toward input-dependent forms of stability and learning dynamics as key topics for future work.

The present results carve out one well-understood corner of theory space: contracting recurrent circuits driven by low-dimensional regular sensory streams. Important open problems include extending these guarantees (a) to learned, non-uniformly contracting networks, (b) to stochastic and chaotic sensory inputs, and (c) to multistable cortical dynamics. One open question is whether predictive-processing and free-energy formulations of cortical inference satisfy analogous contraction conditions, and whether precision-weighted error signals provide a biological mechanism for the stability required here. Nonetheless, even within this restricted regime, our account links dynamical systems theory to empirical neural manifolds and to perceptual phenomena, offering a principled explanation for why low-dimensional sensory manifolds arise and how prediction determines their granularity.

\backmatter

\bmhead{Acknowledgements}
The authors thank their respective universities for their generous support.

\bmhead{Funding}
This research received no specific grant from any funding agency in the public, commercial, or not-for-profit sectors.

\bmhead{Competing interests}
The authors declare no competing interests.

\bmhead{Author contributions}
All authors meet ICMJE criteria for authorship. Conceptualization: V-OS; Formal Analysis: A-Sel; Writing (Original Draft): All authors; Writing (Review \& Editing): All authors.

\bmhead{Generative AI statement}
During preparation of this work the authors used OpenAI ChatGPT, Anthropic Claude, and Google Gemini models to polish written content, assist with the formatting of proofs in \LaTeX{}, and assist with the coding of TikZ figures and R experiments. After using these tools, the authors reviewed and edited the content and take full responsibility for the manuscript and code. Citations were manually retrieved from traditional sources (e.g., PubMed).

\bmhead{Data availability}
This manuscript is theoretical and synthesizes prior results. R code to reproduce all numerical experiments and figures are provided as Supplementary Material.

\begin{appendices}

\section{Glossary of Key Terms}
\label{app:glossary}

This glossary collects definitions of terms used across the three literatures this paper draws on: dynamical systems theory, reservoir computing, and computational neuroscience. Terms are grouped thematically.

\bigskip

\noindent\textbf{Dynamical Systems}

\medskip

\begin{longtable}{@{}>{\raggedright\arraybackslash}p{3.8cm} p{9.2cm}@{}}
\toprule
\textbf{Term} & \textbf{Definition} \\
\midrule
\endfirsthead
\toprule
\textbf{Term} & \textbf{Definition} \\
\midrule
\endhead
\midrule
\multicolumn{2}{r}{\emph{continued on next page}} \\
\bottomrule
\endfoot
\bottomrule
\endlastfoot

Attractor &
A set in state space toward which nearby trajectories converge over time. Examples include fixed points, limit cycles, and strange attractors. \citep{Strogatz2015} \\[6pt]

Bifurcation &
A qualitative change in a system's dynamics (e.g., the appearance or disappearance of an attractor) as a parameter is varied continuously. \citep{Strogatz2015} \\[6pt]

Compact manifold &
A smooth surface (possibly high-dimensional) that is bounded and contains all its limit points. Circles ($S^1$) and tori ($T^k$) are examples. Compactness ensures that continuous functions attain their bounds, which is used throughout the proofs. \citep{Lee2012} \\[6pt]

Conjugacy ($C^r$-conjugacy) &
Two dynamical systems $\phi$ and $\psi$ are $C^r$-conjugate if there exists an invertible, $r$-times differentiable coordinate change $h$ such that $h \circ \phi = \psi \circ h$. Conjugate systems have identical qualitative dynamics; the coordinate change relabels states without altering dynamics. \citep{KatokHasselblatt1995} \\[6pt]

Conjugacy condition number ($\kappa$) &
The product $\|Dh\|_{C^0} \cdot \|Dh^{-1}\|_{C^0}$ measuring how far the coordinate change $h$ is from being distance-preserving. Equals $1$ for rigid rotations; larger values indicate greater metric distortion. Defined in Proposition~\ref{prop:bunching_regular}. \\[6pt]

Delay-coordinate embedding (DCE) &
The reconstruction of a dynamical system's state space from a scalar time series by collecting time-delayed copies of the observed variable into a vector $(\omega(x), \omega(\phi(x)), \ldots, \omega(\phi^{n-1}(x)))$. Under generic conditions, this vector uniquely identifies the system's state. \citep{PackardEtAl1980, Takens1981} \\[6pt]

Diffeomorphism &
A smooth, invertible map whose inverse is also smooth. Used here to describe the driving dynamics $\phi: M \to M$. \citep{Lee2012} \\[6pt]

Embedding &
A smooth, one-to-one map with smooth inverse (on its image). An embedding of $M$ into $\R^N$ means that distinct states remain distinct and the map preserves local geometric structure. \citep{Lee2012} \\[6pt]

Invariant circle / torus &
A circle ($S^1$) or $k$-dimensional torus ($T^k = \R^k/\Z^k$) that is mapped to itself by the dynamics. The regular base systems in this paper live on these objects. Definition~\ref{def:regular_base}. \\[6pt]

Limit cycle &
A closed, isolated periodic orbit that nearby trajectories approach (or recede from) over time. Models periodic phenomena such as oscillations, gait, and rhythmic neural activity. \citep{Strogatz2015} \\[6pt]

Lyapunov exponent &
A quantity measuring the average rate at which nearby trajectories diverge (positive exponent) or converge (negative exponent). Related to the contraction rate $\rho$ in the discrete-time setting. \citep{Strogatz2015} \\[6pt]

Operator norm ($\|\cdot\|_{\mathrm{op}}$) &
See Neural Networks definition. \\[6pt]

Quasiperiodic dynamics &
Motion on a torus with rationally independent frequencies, so that trajectories fill the torus densely without ever exactly repeating. \citep{KatokHasselblatt1995} \\[6pt]

Skew-product &
A coupled system $(x_{t+1}, h_{t+1}) = (\phi(x_t), F(h_t, \omega(x_t)))$ in which the first component (the driver) evolves independently of the second (the driven system). The driver influences the driven system but not vice versa. \citep{stark1999regularity} \\[6pt]

State space &
The set of all possible states of a dynamical system. For a system with $N$ continuous variables, the state space is (a subset of) $\R^N$. Trajectories are paths through state space. \\[6pt]

Strange attractor &
An attractor with sensitive dependence on initial conditions and fractal geometry. Characteristic of chaotic systems. \citep{Strogatz2015} \\[6pt]

Takens' theorem &
For a compact manifold $M$ of dimension $m$, a generic observation function $\omega$, and embedding dimension $n \ge 2m+1$, the delay map $\Phi_\omega^n$ is a $C^1$ embedding. This guarantees that a scalar time series, appropriately lagged, uniquely identifies the system's state. \citep{Takens1981, SauerYorkeCasdagli1991} \\[6pt]

Topological equivalence &
Two spaces or maps are topologically equivalent if they can be related by a continuous, invertible map with continuous inverse (a homeomorphism). Preserves qualitative features such as connectedness and the arrangement of orbits, but not metric properties like distances. \\

Whitney's embedding theorem &
Any compact smooth $m$-dimensional manifold can be smoothly embedded in $\R^{2m+1}$. This sufficient dimension bound is inherited by Takens' theorem and by the generalized synchronization embedding results used here. \citep{Whitney1936} \\[6pt]

\end{longtable}

\bigskip

\noindent\textbf{Reservoir Computing and Generalized Synchronization}

\medskip

\begin{longtable}{@{}>{\raggedright\arraybackslash}p{3.8cm} p{9.2cm}@{}}
\toprule
\textbf{Term} & \textbf{Definition} \\
\midrule
\endfirsthead
\toprule
\textbf{Term} & \textbf{Definition} \\
\midrule
\endhead
\midrule
\multicolumn{2}{r}{\emph{continued on next page}} \\
\bottomrule
\endfoot
\bottomrule
\endlastfoot

Backward expansion &
The rate at which the driving dynamics $\phi$ separates nearby states when traced backward in time, quantified by $\|D\phi^{-1}\|_{C^0(M)}$. Determines how strongly the circuit must contract to achieve a smooth synchronization map. \\[6pt]

Bunching condition &
A sufficient condition for $C^1$ smoothness of the synchronization function. The inequality $\rho \cdot \|D\phi^{-1}\|_{C^0(M)} < 1$ requiring that network contraction dominates backward expansion of the driving dynamics. A standard condition in the theory of normally hyperbolic invariant manifolds and skew-product systems. When satisfied, the synchronization function is $C^1$ smooth, not just continuous. \citep{HirschPughShub1977, stark1999regularity} \\[6pt]

Contraction rate ($\rho$) &
The worst-case factor by which perturbations to the hidden state shrink per time step: $\sup_{(h,u)} \|\partial_h F(h,u)\|_{\mathrm{op}} \le \rho < 1$. Smaller $\rho$ means stronger contraction and faster forgetting of initial conditions. Definition~\ref{def:esp}. \\[6pt]

Echo state map &
Synonym for the synchronization function $f: M \to \R^N$ in the reservoir computing literature. Maps each driving state to the hidden state the network converges to. \citep{Jaeger2001, LukoseviciusJaeger2009} \\[6pt]

Echo state property (ESP) &
The condition that a driven RNN forgets its initial hidden state exponentially fast, so that after transients decay, the hidden state is determined by the driving signal alone. Requires forward invariance of a compact set and uniform contraction for all possible inputs. Definition~\ref{def:esp}. \citep{Jaeger2001, YildizJaegerKiebel2012} \\[6pt]

Fading memory &
The property that a system's current state depends primarily on recent inputs and is progressively less influenced by older inputs. Closely related to ESP; the contraction rate $\rho$ determines the effective memory horizon ($\sim 1/|\log\rho|$ steps; e.g., $\rho$ = 0.95 gives an effective memory of about 20 steps). \citep{BoydChua1985, Jaeger2001} \\[6pt]

Generalized synchronization (GS) &
The regime in which a driven system's long-run state becomes a deterministic function of the driving state. Defined by the existence of a synchronization function $f$ satisfying an invariance condition and an attraction condition. Definition~\ref{def:gs}. \citep{Rulkov1995, stark1999regularity, HartHookDawes2020} \\[6pt]

Generic &
A property that holds for ``most'' choices of parameters or functions, in the precise sense of holding on a residual (countable intersection of open dense) subset of the relevant function space. Embedding theorems guarantee that generic observation functions yield embeddings. \citep{Takens1981, Hart2025} \\[6pt]

Input-dependent ESP &
A relaxation of ESP in which the contraction condition holds only for the structured inputs the network actually receives, rather than for all possible inputs. More biologically realistic for sensory circuits. \citep{ManjunathJaeger2013, YildizJaegerKiebel2012} \\[6pt]

Prevalence &
A measure-theoretic notion of ``almost every'' for infinite-dimensional function spaces, on which no Lebesgue measure exists. A property is prevalent if it fails only on a \emph{shy} set, the analogue of a set of measure zero \citep{HuntSauerYorke1992}. Brought to delay-coordinate (Takens) embedding by \citet{SauerYorkeCasdagli1991}. \\[6pt]

Reservoir computing &
A computational framework in which a fixed (untrained) recurrent network (the reservoir) is driven by input, and only a readout layer is trained. The reservoir's dynamics implicitly perform delay-coordinate embedding. \citep{Jaeger2001, MaassNatschlaegerMarkram2002, LukoseviciusJaeger2009} \\[6pt]

Separation modulus &
A function $\eta(\delta)$ quantifying the minimum output difference guaranteed when inputs are at least $\delta$ apart: $d_M(x,x') \ge \delta \implies \|g(x)-g(x')\| \ge \eta(\delta)$. Any continuous injective map on a compact domain admits a positive separation modulus. Used in Proposition~\ref{prop:prediction_separation} to convert prediction error bounds into representational resolution limits. \\[6pt]

Synchronization function ($f$) &
The map $f: M \to \R^N$ sending each state of the driving system to the unique hidden state the network converges to when driven from that state. The central mathematical object of this paper. Definition~\ref{def:gs}, Theorem~\ref{thm:gs_existence}. \\[6pt]

Weak generalized synchronization &
A regime in which $f$ is continuous but not differentiable, arising when backward expansion dominates contraction. Does not occur in the regular regime treated here under the sufficient conditions of Proposition~\ref{prop:no_weak_gs}. \citep{KellerJafriRamaswamy2013} \\

\end{longtable}

\bigskip

\noindent\textbf{Neural Networks}

\medskip

\begin{longtable}{@{}>{\raggedright\arraybackslash}p{3.8cm} p{9.2cm}@{}}
\toprule
\textbf{Term} & \textbf{Definition} \\
\midrule
\endfirsthead
\toprule
\textbf{Term} & \textbf{Definition} \\
\midrule
\endhead
\midrule
\multicolumn{2}{r}{\emph{continued on next page}} \\
\bottomrule
\endfoot
\bottomrule
\endlastfoot

Gated recurrent unit (GRU) &
A recurrent architecture that uses learned gates to control how much of the previous hidden state is retained versus updated. Similar in function to LSTM but with fewer parameters. \citep{ChoEtAl2014} \\[6pt]

Hidden state ($h_t \in \R^N$) &
The internal state of a recurrent neural network at time $t$, consisting of $N$ real-valued activations. Evolves according to the state update $h_{t+1} = F(h_t, u_t)$. \\[6pt]

Long short-term memory (LSTM) &
A recurrent neural network architecture with gated memory cells that can maintain information over long time intervals, mitigating the vanishing gradient problem. \citep{HochreiterSchmidhuber1997} \\[6pt]

Recurrent neural network (RNN) &
A neural network containing feedback connections, so that the hidden state at time $t+1$ depends on both the current input and the previous hidden state. This recurrence gives the network memory of past inputs. \\[6pt]

Spectral radius &
The largest absolute eigenvalue of the weight matrix $W$. A rough indicator of network stability: spectral radius less than 1 is necessary (but not sufficient) for ESP in linear networks. \citep{Jaeger2001} \\[6pt]

State-space model (SSM) &
A neural architecture that processes sequences through a learned linear dynamical system, updated at each time step. Recent examples include structured state-space layers (S4, Mamba). SSMs have structural similarities to the driven RNN model used here. \citep{GuEtAl2022, GuDao2024} \\[6pt]

Transformer &
A neural architecture based on attention mechanisms rather than recurrence. Processes all positions in a sequence simultaneously. Lacks the intrinsic temporal recurrence central to the present analysis. \citep{VaswaniEtAl2017} \\[6pt]

Truncated backpropagation through time (BPTT) &
A training algorithm for recurrent networks. The loss gradient is computed by unrolling the network's recurrence for a fixed number of time steps and applying the chain rule backward through those steps. ``Truncated'' means the gradient computation is cut off after a fixed window rather than extending to the beginning of the input sequence. \\[6pt]

Operator norm ($\|W\|_{\mathrm{op}}$) &
The largest singular value of the weight matrix $W$. Measures the worst-case factor by which $W$ can stretch a vector. Distinct from the spectral radius (largest absolute eigenvalue): for non-normal matrices, the operator norm can exceed the spectral radius. A sufficient condition for the echo state property in $\tanh$ RNNs is $\|W\|_{\mathrm{op}} < 1$. \\[6pt]

Ridge regression &
A linear regression method that adds a penalty proportional to the squared magnitude of the coefficients: $\hat{\beta} = (X^\top X + \lambda I)^{-1} X^\top y$, where $\lambda > 0$ is a regularization parameter. Used here for the readout layer in Experiment 3 because it is deterministic (no training noise) and has a closed-form solution. \\[6pt]

Collision metric &
Related to the injectivity condition of the synchronization map: zero collisions implies the map separates all sampled states. A measure of embedding quality used in the numerical experiments. For pairs of states that are well separated on the base manifold, the collision fraction is the proportion whose hidden-state representations are nearly coincident. Low collision fraction indicates that the synchronization map preserves state separation. \\[6pt]

Principal component analysis (PCA) &
A dimensionality reduction method that finds the orthogonal directions of greatest variance in a dataset. Used here to project high-dimensional hidden states into two or three dimensions for visualization. PCA is linear and can miss curved manifold structure, but suffices when the manifold is low-dimensional relative to the hidden space. \\[6pt]

Persistent homology &
A method from topological data analysis that detects topological features (connected components, loops, voids) in a point cloud at multiple spatial scales. Each feature is born at one scale and dies at another; long-lived features indicate genuine structure rather than noise. A circle produces one long-lived $H_1$ bar. A torus produces two long-lived $H_1$ bars and one long-lived $H_2$ bar. In practice, finite sampling and projection artifacts produce additional short-lived bars (see Experiment 4). \citep{Edelsbrunner2010} \\

\end{longtable}

\bigskip

\noindent\textbf{Neural Coding and Population Dynamics}

\medskip

\begin{longtable}{@{}>{\raggedright\arraybackslash}p{3.8cm} p{9.2cm}@{}}
\toprule
\textbf{Term} & \textbf{Definition} \\
\midrule
\endfirsthead
\toprule
\textbf{Term} & \textbf{Definition} \\
\midrule
\endhead
\midrule
\multicolumn{2}{r}{\emph{continued on next page}} \\
\bottomrule
\endfoot
\bottomrule
\endlastfoot

Computation through dynamics &
The framework that neural computation consists of the temporal evolution of population state through a dynamical system, rather than static encoding of variables in individual neuron firing rates. \citep{VyasEtAl2020} \\[6pt]

Intrinsic vs.\ embedding dimensionality &
Intrinsic dimensionality is the number of independent latent variables underlying population activity; embedding dimensionality is the number of linear dimensions (e.g., principal components) needed to represent the manifold in neural state space. A ring has intrinsic dimension~1 but may require many embedding dimensions because it is curved. \citep{JazayeriOstojic2021} \\[6pt]

Neural manifold hypothesis &
The proposal that task-relevant neural population activity is confined to a low-dimensional subspace (the manifold) embedded in the high-dimensional space of all possible activity patterns. Supported by empirical findings in motor cortex, visual cortex, and navigation circuits. \citep{CunninghamYu2014, Gallego2017} \\[6pt]

Ring attractor &
A neural circuit whose population activity forms a persistent bump on a one-dimensional circular manifold, encoding a continuous circular variable such as heading direction. Confirmed experimentally in \emph{Drosophila} \citep{Kim2017} and mouse thalamus \citep{Chaudhuri2019}. \\[6pt]

\end{longtable}

\bigskip

\noindent\textbf{Computational Neuroscience}

\medskip

\begin{longtable}{@{}>{\raggedright\arraybackslash}p{3.8cm} p{9.2cm}@{}}
\toprule
\textbf{Term} & \textbf{Definition} \\
\midrule
\endfirsthead
\toprule
\textbf{Term} & \textbf{Definition} \\
\midrule
\endhead
\midrule
\multicolumn{2}{r}{\emph{continued on next page}} \\
\bottomrule
\endfoot
\bottomrule
\endlastfoot

Categorical perception &
The phenomenon whereby a continuum of physical stimuli is perceived as falling into discrete categories, with better discrimination across category boundaries than within them. \citep{Harnad1987} \\[6pt]

Discrimination threshold &
The minimum physical difference between two stimuli that a perceiver can reliably detect. In the present framework, set by the finite prediction error $\varepsilon$. \\[6pt]

Metameric collapse &
The perceptual identification of physically distinct stimuli that produce indistinguishable neural responses. In color vision, metamers are different spectral distributions that appear identical. In the present framework, arises when distinct driving states have near-identical futures. \citep{Wandell1995} \\[6pt]

Neural manifold &
The low-dimensional surface in neural state space along which population activity is concentrated during a task. Empirically estimated via dimensionality reduction methods. \citep{ChurchlandEtAl2012, Gallego2017} \\[6pt]

Neural population state &
The vector of firing rates or activity levels across a population of neurons at a given time, treated as a point in $\R^N$. The biological counterpart of the hidden state $h_t$. \\[6pt]

Opponent-process coding &
The encoding of sensory information as the difference between opposing channels. In color vision: L--M (red--green) and S--(L+M) (blue--yellow) channels. Produces a low-dimensional representation of chromatic information. \citep{Wandell1995} \\[6pt]

Predictive processing &
The theoretical framework in which the brain continuously generates predictions about incoming sensory signals and updates its internal model based on prediction errors. \citep{Friston2010, clark2013} \\[6pt]

Quality space &
The proposal that each experience corresponds to a point in a multidimensional space defined over the activity of processing units, so that similarity of experience reflects proximity in representational geometry \citep{Dolega2025-bl}. Related to the neural manifold hypothesis; the present framework provides dynamical conditions under which such spaces arise from sensory-driven recurrent dynamics. \\[6pt]

Variability quenching &
The reduction in trial-to-trial variability of neural responses following stimulus onset. Evidence that sensory input drives the network toward a reproducible state, consistent with the echo state property. \citep{HennequinEtAl2018} \\

\end{longtable}

\section{Numerical Experiment Methods and Full Results}
\label{app:experiments}

This appendix provides complete methodological details for the four numerical experiments summarized in Section~\ref{subsec:numerical_experiments}. All code is written in R and will be made available in a public repository upon publication.

\subsection{Software and Reproducibility}

All experiments were run in R version 4.4.1 \citep{RCoreTeam2024} on Windows 11 (x86\_64). The following packages were used:
\begin{itemize}[nosep]
\item \texttt{ggplot2} (version 3.5.1) for all figures \citep{Wickham2016}
\item \texttt{gridExtra} (version 2.3) for multi-panel figure layout
\item \texttt{TDAstats} (version 0.4.1) for persistent homology computations \citep{Wadhwa2018}
\end{itemize}
Random seeds are set at the beginning of the script (\texttt{set.seed(42)} for the observation function parameters, \texttt{set.seed(123)} for training) and at the start of each replicate run. The observation functions (cosine tuning curves for the circle driver, sine sums for the torus driver) use fixed parameters across all experiments so that results are comparable.

\subsection{Network Architecture}

All experiments use a single-layer $\tanh$ recurrent neural network:
\begin{align}
h_{t+1} &= \tanh(W h_t + W_{\mathrm{in}} u_t + b), \\
\hat{u}_{t+1} &= w_{\mathrm{out}}^\top h_{t+1} + c_{\mathrm{out}},
\end{align}
where $h_t \in \R^N$ is the hidden state, $u_t \in \R$ is the scalar input, $W \in \R^{N \times N}$ is the recurrent weight matrix, $W_{\mathrm{in}} \in \R^{N \times 1}$ is the input weight vector, $b \in \R^N$ is a bias vector, $w_{\mathrm{out}} \in \R^{1 \times N}$ is the readout weight vector, and $c_{\mathrm{out}} \in \R$ is the readout bias. The nonlinearity is applied elementwise.

\paragraph{Initialization.} The recurrent weight matrix is initialized from a Gaussian distribution and then rescaled so that its operator norm (largest singular value) equals a prescribed spectral scale $s$:
\[
W_{\mathrm{raw}} \sim \mathcal{N}(0, 1/N), \qquad W = \frac{s}{\sigma_{\max}(W_{\mathrm{raw}})} W_{\mathrm{raw}},
\]
where $\sigma_{\max}$ denotes the largest singular value. This guarantees $\|W\|_{\mathrm{op}} = s$ at initialization. Input weights are drawn from $\mathcal{N}(0, 0.25)$, biases are initialized to zero, and readout weights are drawn from $\mathcal{N}(0, 1/N)$.

\paragraph{Training.} Networks are trained to minimize one-step prediction error:
\[
\mathcal{L} = \frac{1}{L} \sum_{t=1}^{L} (\hat{u}_{t+1} - u_{t+1})^2,
\]
on subsequences of length $L = 200$ drawn randomly from the full trajectory ($T_{\mathrm{total}} = 8000$--$12000$ depending on experiment). The Adam optimizer is used with learning rate $10^{-3}$, $\beta_1 = 0.9$, $\beta_2 = 0.999$, $\epsilon_{\mathrm{Adam}} = 10^{-8}$, and batch size 10 (10 randomly drawn subsequences per gradient update). Gradients are clipped to a maximum norm of 5.0 to prevent numerical instability. The hidden state is initialized to zero at the start of each training subsequence.

\paragraph{Post-training verification.} After training, we recompute the operator norm $\|W\|_{\mathrm{op}}$ and spectral radius $\max_i |\lambda_i(W)|$ of the trained recurrent weight matrix. If the operator norm exceeds 1, the strict echo state property is no longer guaranteed. We report these values for every trained network.

\paragraph{Train--test considerations.} Only Experiment~3 uses a held-out test set, because it evaluates prediction accuracy. Experiments~1, 2, and 4 evaluate geometric and dynamical properties of the learned representation (GS convergence, collision rates, manifold topology) rather than prediction accuracy on held-out data. Overfitting to the training signal would manifest as failure of GS convergence or degraded manifold structure, neither of which was observed.

\subsection{Driving Dynamics}

\paragraph{Circle driver ($d=1$).} Quasiperiodic rotation on $S^1$:
\[
\theta_{t+1} = (\theta_t + \alpha) \bmod 2\pi, \qquad \alpha = 2\pi(\sqrt{2} - 1).
\]
The observation is a sum of cosine tuning curves with fixed preferred directions and amplitudes:
\[
\omega(\theta) = \sum_{j=1}^{6} a_j \cos(\theta - \mu_j),
\]
where $\mu_j$ are equally spaced on $[0, 2\pi)$ and amplitudes $a_j \in [0.5, 1.5]$ are drawn once and reused across all experiments. The scalar input $u_t = \tilde{\omega}(\theta_t)$ is the z-scored version (zero mean, unit variance).

\paragraph{Logistic driver (chaotic).} The logistic map at $r = 3.9$:
\[
x_{t+1} = 3.9 \, x_t (1 - x_t), \qquad x_1 \sim \mathcal{U}(0,1).
\]
The scalar input is the z-scored $x_t$.

\paragraph{Torus driver ($d=2$).} Quasiperiodic translation on $T^2$:
\[
\bm{\theta}_{t+1} = (\bm{\theta}_t + \bm{\alpha}) \bmod 2\pi, \qquad \bm{\alpha} = 2\pi(1/\sqrt{2},\; 1/\sqrt{3}).
\]
The observation is $\omega(\theta_1, \theta_2) = a_1 \sin\theta_1 + a_2 \sin\theta_2$, where $a_1, a_2$ are drawn once from $\mathcal{U}(0.8, 1.2)$ and reused across all experiments. The scalar input is z-scored.

\subsection{Experiment 1: Generalized Synchronization}

\paragraph{Design.} Four conditions: \{circle, logistic\} $\times$ \{$s=0.5$, $s=0.95$\}. All use $N=8$, trained for 200 epochs.

\paragraph{GS assessment.} After training, the network is driven by the first 2000 time steps of the input from five random initial hidden states $h_0^{(i)} \sim \mathcal{N}(0, I)$ and one reference initial state $h_0^{\mathrm{ref}} = 0$. At each time step, we compute $d_t^{(i)} = \|h_t^{(i)} - h_t^{\mathrm{ref}}\|_2$. Convergence of all $d_t^{(i)}$ to zero indicates generalized synchronization. We also project the hidden trajectories into the first two principal components of the reference trajectory to visualize convergence in state space.

\paragraph{Results.} Post-training operator norms: circle-strong 1.21, circle-weak 1.06, logistic-strong 1.03, logistic-weak 1.36. All exceed 1, so strict ESP is not maintained after training. GS convergence is observed in all four conditions, with the fastest convergence for regular driving with strong initial contraction and the slowest for chaotic driving with weak initial contraction.

\subsection{Experiment 2: Embedding Dimension}

\paragraph{Design.} Circle ($d=1$): $N \in \{2,3,4,6,8\}$. Torus ($d=2$): $N \in \{3,4,5,8,12\}$. Spectral scale $s=0.8$, trained for 100 epochs. Three random seeds per condition.

\paragraph{Collision metric.} After training, the network is driven for $T_{\mathrm{vis}} = 4000$ steps from $h_0 = 0$. We sample 5000 random index pairs $(i_1, i_2)$ and compute the base-manifold distance (wrapped angular distance for the circle; Euclidean distance of wrapped component differences for the torus) and the hidden-space Euclidean distance. Pairs with base distance $\ge \delta_x$ ($\delta_x = 0.2$ for circle, $0.4$ for torus) are classified as ``distinct.'' Among distinct pairs, the collision fraction is the proportion with hidden distance $< \delta_h = 0.1$.

\paragraph{Results.} Collision fractions decrease with $N$, with the steepest reduction near $N = 2d+1$. Most trained networks maintain $\|W\|_{\mathrm{op}} < 1$ after training, likely reflecting the moderate spectral scale ($s=0.8$) and shorter training duration (100 epochs) relative to the other experiments.

\subsection{Experiment 3: Prediction-Separation}

\paragraph{Design.} Three RNNs ($N=8$) trained on the same circle driver with $s \in \{0.3, 0.7, 0.95\}$, 150 epochs each. A linear readout $P:\R^N \to \R^{K+1}$ (here $\R^8 \to \R^6$) is trained via ridge regression ($\lambda = 10^{-3}$) on the first half of the trajectory to predict $K=5$ future observations from the hidden state, and evaluated on the second half.

\paragraph{Prediction error.} For each test time point, the prediction error is $\varepsilon_t = \|P(h_t) - \Gamma_K(x_t)\|_2$, where $P$ is the trained readout and $\Gamma_K$ is the $K$-step forward map. We report the mean and supremum over the test set.

\paragraph{State separation.} We sample 5000 pairs of test states and compute three quantities: base-manifold distance $d_M$, future difference $\|\Gamma_K(x) - \Gamma_K(x')\|$, and hidden distance $\|h(x) - h(x')\|$.

\paragraph{Results.} Mean prediction errors: 0.065 (strong), 0.036 (moderate), 0.029 (weak). Sample suprema: 0.175, 0.090, 0.057, giving supremum-to-mean ratios 2.70, 2.49, 1.99 (computed from the unrounded errors). Post-training operator norms: 0.72 (strong, ESP holds), 1.27 (moderate, ESP not guaranteed), 1.48 (weak, ESP not guaranteed). Participation ratios of the hidden-state covariance: 1.19, 1.34, 1.62 (of $N=8$). The less contractive networks achieve lower prediction error, possibly reflecting greater representational capacity. The coupling between future differences and hidden-state separation is visible across all three regimes, consistent with Proposition~\ref{prop:prediction_separation}.

\subsection{Experiment 4: Manifold Topology}

\paragraph{Design.} $S^1$ driver: $N=8$, $s=0.8$, 150 epochs. $T^2$ driver: $N=12$, $s=0.8$, 200 epochs. After training, hidden states are collected over 4000 steps ($S^1$) or 6000 steps ($T^2$) from $h_0 = 0$.

\paragraph{PCA.} Principal component analysis is applied to the hidden-state matrix (centered, not scaled). For $S^1$, the first three PCs capture 99.8\% of variance (76.6\%, 22.9\%, 0.4\%). For $T^2$, the first four PCs capture 99.6\% (67.8\%, 25.4\%, 5.5\%, 0.8\%).

\paragraph{Persistent homology.} Two protocols are applied:
\begin{enumerate}[nosep]
\item \emph{Point-wise}: 300 hidden states are subsampled randomly (ignoring temporal order). The pairwise Euclidean distance matrix is computed on the first 3 PCs ($S^1$) or 4 PCs ($T^2$), and Vietoris-Rips persistent homology is computed up to homological dimension 1 ($S^1$) or 2 ($T^2$), with the maximum filtration radius set automatically by \texttt{TDAstats} as the diameter of the point cloud.
\item \emph{Sequential}: The first 300 consecutive hidden states are taken in trajectory order. The same PH computation is applied to their pairwise distance matrix.
\end{enumerate}

\paragraph{Results.} For $S^1$, both protocols detect a single dominant $H_1$ bar with infinite persistence gap (ratio of longest to second-longest bar), confirming ring topology. For $T^2$, point-wise PH detects 67 $H_1$ bars and 8 $H_2$ bars; sequential PH detects 60 $H_1$ bars and 8 $H_2$ bars. The longest-to-second-longest gap ratios are 3.09 ($H_1$, point-wise), 1.09 ($H_2$, point-wise), 4.64 ($H_1$, sequential), and 1.01 ($H_2$, sequential). Against a 95th-percentile surrogate null (Gaussian point clouds matched in size, dimension, and scale for the point-wise protocol; per-coordinate Fourier phase-randomized series for the sequential protocol, 50 surrogates each), the point-wise protocol has two $H_1$ bars and zero $H_2$ bars above the null, and the sequential protocol has one $H_1$ and zero $H_2$. The two point-wise $H_1$ bars match the two independent cycles of a torus; the $H_2$ void is not separated from noise at 300 points in a four-dimensional projection.

\paragraph{Post-training contraction.} $S^1$ network: $\|W\|_{\mathrm{op}} = 1.06$, spectral radius $= 0.68$. $T^2$ network: $\|W\|_{\mathrm{op}} = 1.20$, spectral radius $= 1.01$. Both exceed strict ESP after training.

\subsection{Additional Diagnostics for Experiments 1, 3, and 4}
\paragraph{Conditional Lyapunov exponent.} For each trained network of Experiment~1 we form the fiber Jacobian along the driven orbit, $J_t = \operatorname{diag}(1 - h_{t+1}^2)\,W$, and estimate the top conditional Lyapunov exponent $\lambda_{\mathrm{cond}} = \lim_{T\to\infty}\tfrac1T\log\|J_{T-1}\cdots J_0\|$ by power iteration with renormalization over 3000 steps after a 300-step burn-in, also reporting the mean and maximum operator norm of $J_t$. Input-dependent contraction is corroborated by driving each network from several initial hidden states under the trained signal and under an arbitrary standardized white-noise signal and measuring convergence to the reference trajectory.

\paragraph{Participation ratio.} For each Experiment~3 regime, $\mathrm{PR} = (\sum_i \lambda_i)^2 / \sum_i \lambda_i^2$, where $\lambda_i$ are the eigenvalues of the hidden-state covariance.

\paragraph{Persistent-homology surrogate null.} For each protocol we generate 50 surrogates and take the 95th percentile of the maximum surrogate persistence per homology dimension as the threshold. Point-wise surrogates are Gaussian point clouds matched in sample size, dimension, and per-coordinate mean and standard deviation. Sequential surrogates are per-coordinate Fourier phase-randomized series, which preserve each coordinate's power spectrum while destroying deterministic loop closure.

\par

\subsection{Experiment 5: Scaling to Large Hidden Dimension}
\paragraph{Methods.} Experiment~5 repeats the Experiment~2 collision metric at $N\in\{50,100,200\}$ for both drivers, with a single seed and a reduced training budget (40 epochs, sequence length 120, batch size 6).

\begin{figure}[ht]
\centering
\includegraphics[width=0.6\textwidth]{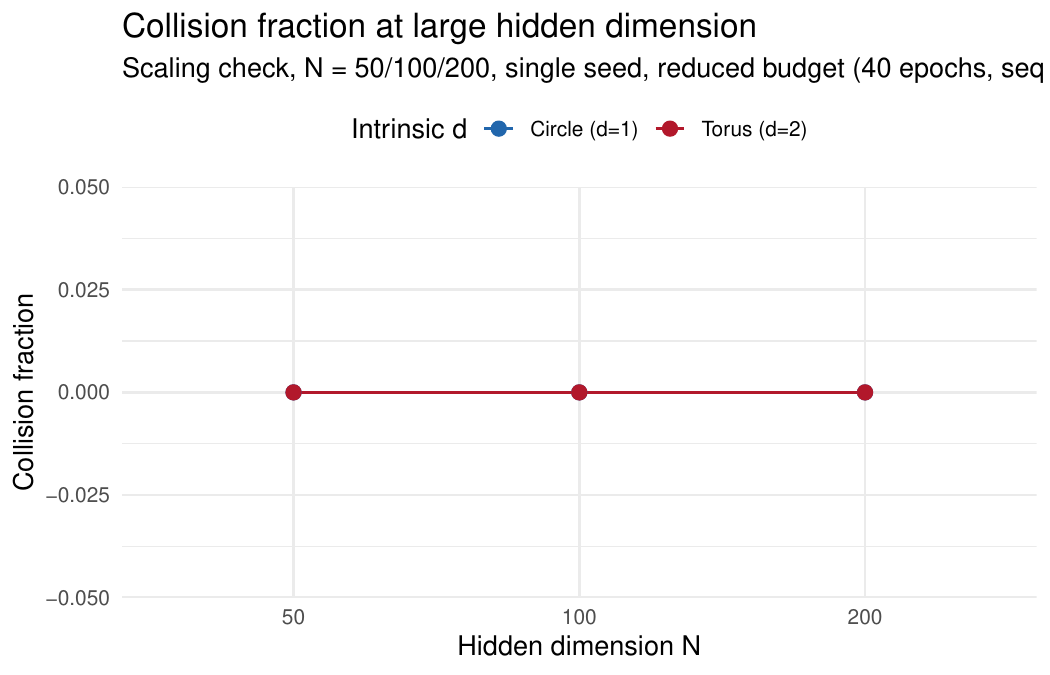}
\caption{Collision fraction at large hidden dimension. For the circle ($d=1$) and torus ($d=2$) drivers the collision fraction is zero at $N=50,100,200$ (single seed, reduced training budget). Embedding quality does not degrade as the hidden dimension grows.}
\label{fig:largeN}

\end{figure}

Neural population manifolds are recorded from hundreds to thousands of neurons, so we check that the embedding behavior is not an artifact of the small-$N$ regime. We repeated the collision-metric analysis of Experiment~2 at $N\in\{50,100,200\}$ for the circle ($d=1$) and torus ($d=2$) drivers. As a scaling check, this sweep uses a single seed and a reduced training budget (40 epochs, sequence length 120, batch size 6). The collision fraction was $0$ at every $N$ for both drivers (Figure~\ref{fig:largeN}), with post-training operator norms between $0.91$ and $1.37$. Embedding quality persists as $N$ grows well beyond the threshold; the small-$N$ results lie on the same trend. Here $N$ is the embedding dimension while the intrinsic dimension $d$ is unchanged \citep{JazayeriOstojic2021}.
\par

\subsection{Experiment 6: Embedding Without Training}
\paragraph{Methods.} The frozen reservoir uses the same initialization as the trained networks (recurrent weight matrix scaled to a prescribed operator norm, input weights $\sim\mathcal{N}(0,0.5^2)$, zero bias) with the recurrent and input weights left untrained; the readout is unused and set to zero. We sweep the Experiment~2 hidden dimensions at operator norm $0.8$ (matched to the trained networks) and $0.9$, three random reservoir draws each, and compare against trained networks reproduced from the Experiment~2 seeds. Collision pairs and base-distance thresholds are those of Experiment~2; the scale-normalized collision divides each sampled hidden distance by the cloud radius, the root-mean-square distance of all hidden states from their mean (computed over the full evaluation window, while collisions are taken over $5000$ sampled pairs). Each network is normalized by its own cloud radius. The co-evolution panel trains one network (circle, $N=8$) and records, at epochs $\{1,5,10,20,40,80,150\}$ from a fixed seed, the recurrent-weight movement $\|W_e-W_0\|_F$ and the normalized collision fraction.

Every experiment so far trains the recurrent weights. To separate what the driving dynamics contribute from what training adds, we ask whether a network embeds its input with the recurrent weights frozen. We build a random reservoir with operator norm $\|W\|_{\mathrm{op}}=0.8$, so the strict echo state property holds by construction, leave the recurrent and input weights untrained, and drive it with the same circle and torus signals. The collision metric of Experiment~2 is made scale-invariant by dividing hidden distances by the cloud radius, the root-mean-square distance of hidden states from their mean, so a frozen reservoir and a trained network are compared on the same footing.

\begin{figure}[ht]
\centering
\includegraphics[width=0.9\textwidth]{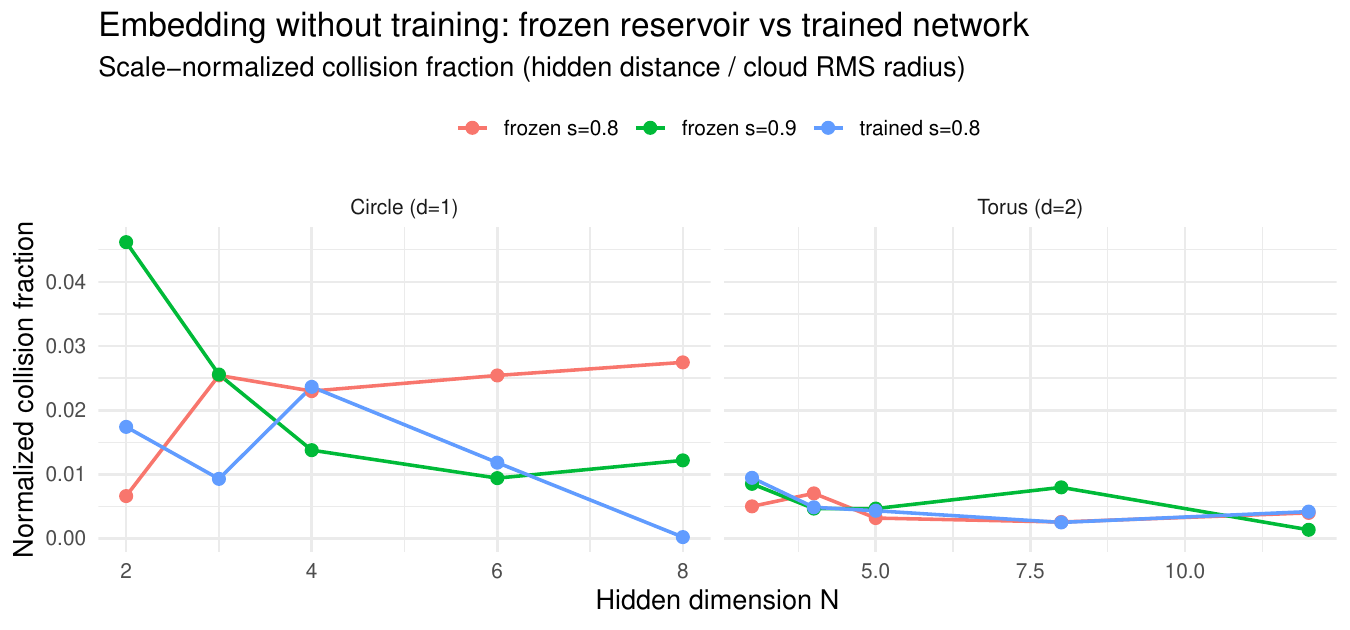}
\caption{Embedding without training. Scale-normalized collision fraction (hidden distance divided by cloud radius) against hidden dimension $N$ for a frozen reservoir (operator norm $0.8$ and $0.9$, recurrent weights untrained) and a trained network ($0.8$ initialization), for the circle ($d=1$) and torus ($d=2$) drivers. The frozen reservoir embeds at every $N$; on the torus it matches the trained network, and on the circle training lowers collisions further at large $N$. The cloud radius grows with $N$, so values are compared between arms at fixed $N$.}
\label{fig:reservoir}

\end{figure}

The frozen reservoir embeds. Generalized synchronization holds (trajectories from different initial hidden states converge, residual divergence zero) and the top conditional Lyapunov exponent is negative at every hidden dimension (circle $-0.41$ to $-0.95$, torus $-1.31$ to $-0.68$). Normalized collision fractions are low throughout: for the operator-norm-$0.8$ reservoir, below $0.03$ for the circle and below $0.008$ for the torus, and the strict-echo-state $0.9$ reservoir is similar (Figure~\ref{fig:reservoir}). On the torus the frozen reservoir matches the trained network (both near $0.003$--$0.008$), while on the circle training lowers collisions further at large hidden dimension, reaching $0.0002$ at $N=8$ against the reservoir's $0.027$. The recurrent weights are therefore not required to produce the embedding; training only refines it. Because the cloud radius grows with $N$, the normalized collision is compared across arms at fixed $N$ rather than read as a threshold curve.

\begin{figure}[ht]
\centering
\includegraphics[width=0.7\textwidth]{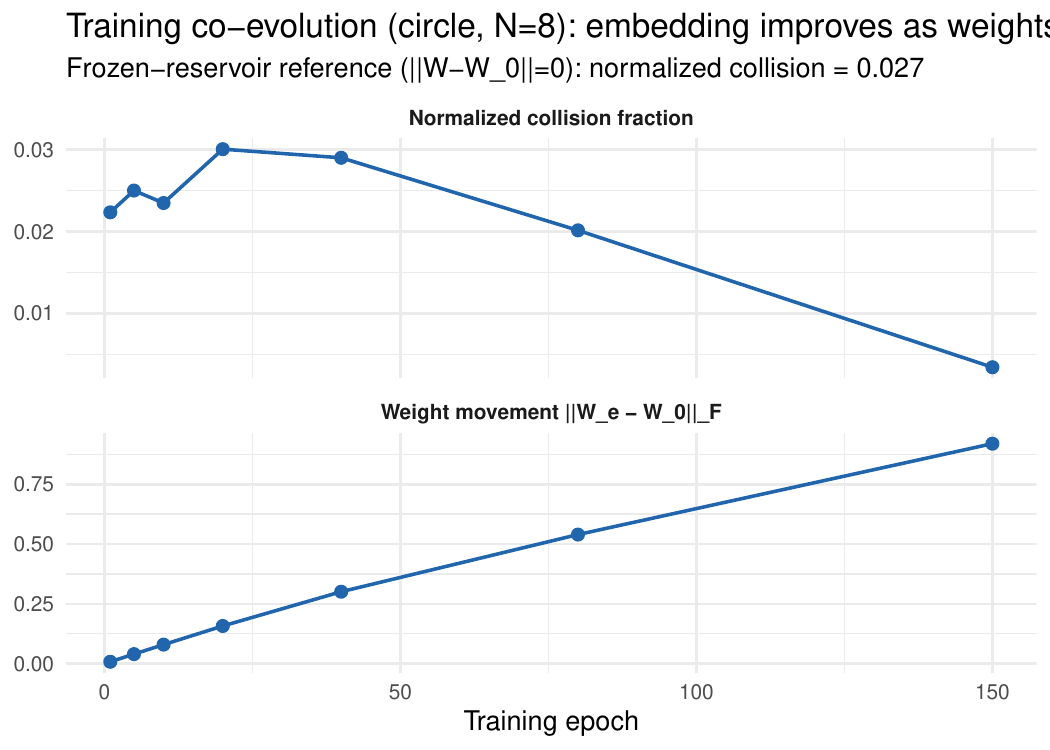}
\caption{Training co-evolution (circle, $N=8$). Top: the normalized collision fraction. Bottom: the recurrent-weight movement $\|W_e-W_0\|_F$ from initialization. As the weights move, the embedding improves, reaching $0.003$ by epoch $150$; the frozen reservoir (zero weight movement) sits at $0.027$.}
\label{fig:coevolution}

\end{figure}

Figure~\ref{fig:coevolution} tracks this refinement during training. As one network trains, its recurrent weights move away from initialization, with $\|W_e-W_0\|_F$ growing from $0.008$ after one epoch to $0.92$ after $150$, and its embedding improves: the normalized collision fraction, comparable to the frozen reservoir's $0.027$ through mid-training, falls to $0.003$ by the end. The reservoir is the zero-movement reference. The weights and the realized representation move together toward a cleaner embedding.
\par

\subsection{Experiment 7: Novel Trajectories and Recovery After a Different Attractor}
\paragraph{Methods.} A novel orbit is a fresh quasiperiodic realization (new random phase) of length $4000$, realized once and reused for all conditions; the chaotic displacer is a separate logistic realization ($r=3.9$) of $500$ steps. For each network (trained and frozen, circle $N=8$ and torus $N=12$) the reference trajectory is driven from the zero state; the immediate trajectory from a random initial hidden state; the cross-attractor trajectory from the hidden state reached after the $500$ logistic steps, then driven by the same novel orbit. Divergence at step $k$ compares the same novel-orbit input. The re-synchronization step is the first after which divergence stays below $0.05$ times the cloud radius (a per-network threshold, since the radius varies by network) through the final tenth of the run (right-censored to undefined otherwise). The recovered embedding is the normalized collision fraction on the post-transient hidden states (first $500$ steps discarded), each regime normalized by its own post-transient cloud radius.

The synchronization function maps each environmental state to a hidden state, so the embedding should hold on any orbit of the same dynamics and should not depend on where the hidden state has been. We test both. We generate a novel orbit, a fresh quasiperiodic realization the network never saw; because the orbit is dense, this is a new trajectory on the same manifold, not a new region of it. We drive each network with this orbit from three histories: from the zero state (the reference trajectory), from a random initial hidden state, and after first driving the hidden state for $500$ steps on the logistic chaotic map (Section~\ref{subsec:numerical_experiments}) and then switching to the novel orbit.

\begin{figure}[ht]
\centering
\includegraphics[width=0.9\textwidth]{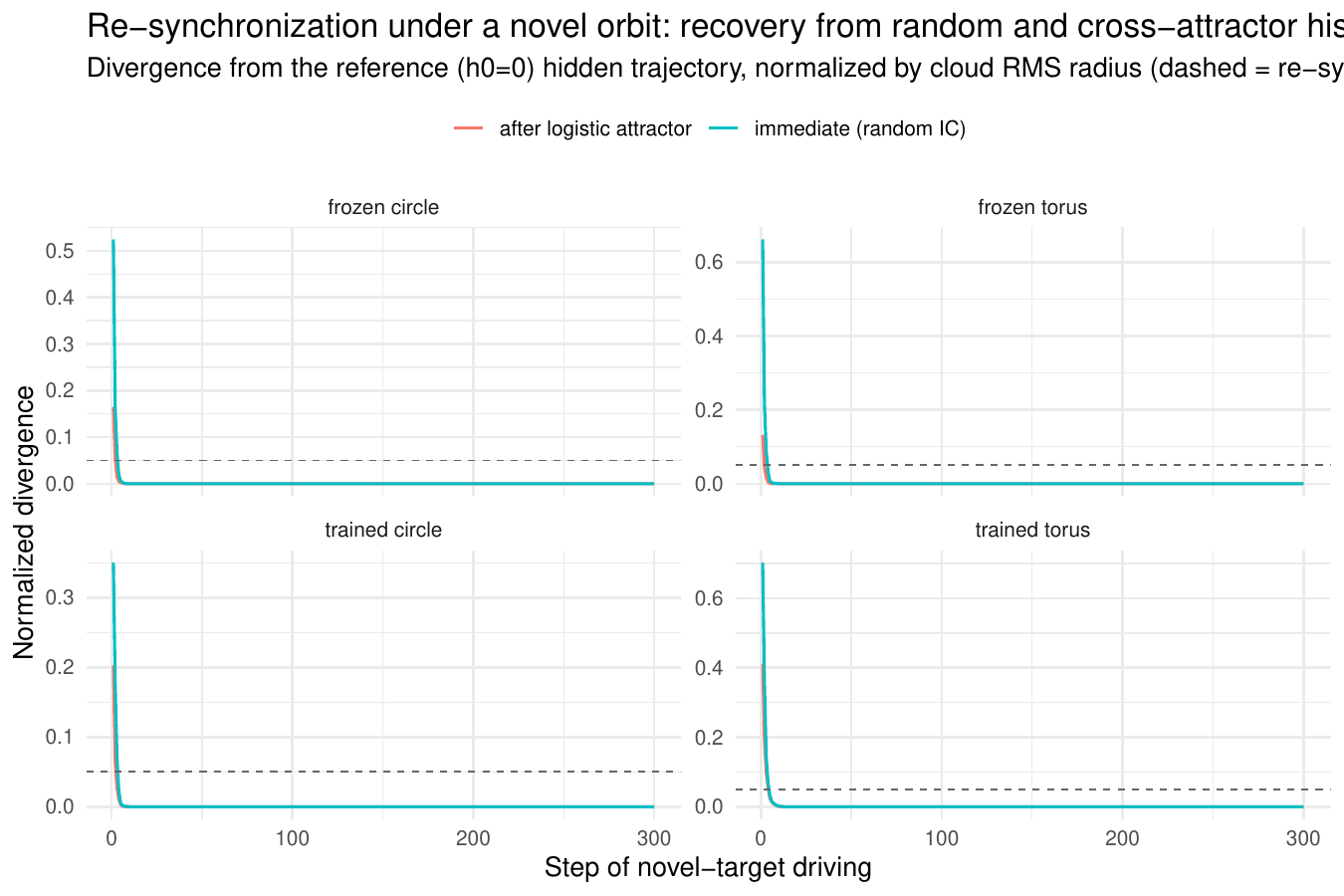}
\caption{Recovery to the same hidden trajectory under a novel orbit. Divergence from the reference (zero-state) trajectory, normalized by the cloud radius, for a trajectory started from a random hidden state and for one driven first on the logistic chaotic map and then switched to the novel orbit. Both collapse to the reference within a few steps for all four networks; the dashed line marks the re-synchronization threshold.}
\label{fig:recovery}

\end{figure}

The embedding holds on the novel orbit (normalized collision $0.0002$ trained circle, $0.020$ frozen circle, $0.003$--$0.004$ torus), matching the values on the training orbit. From both the random and the cross-attractor histories the network re-synchronizes to the reference trajectory within two to five steps (Figure~\ref{fig:recovery}); the starting states are genuinely displaced, with initial divergence between $0.13$ and $0.70$ cloud radii, well above the re-synchronization threshold. After the transient the trajectories coincide: divergence from the reference falls to zero, and the recovered embedding matches the reference value regardless of history. The embedding depends on the current driving signal, not on the hidden state's past. This realizes the echo state property of Section~\ref{sec:preliminaries}: the network forgets a chaotic excursion and re-enters the same internal manifold once the regular signal returns.
\par

\subsection{Experiment 8: Training Redistributes the Contraction Spectrum}
\paragraph{Methods.} The full spectrum of the fiber-Jacobian cocycle $J_t=\operatorname{diag}(1-h_{t+1}^2)W$ along the driven orbit is computed by a QR (Benettin) iteration: an orthonormal frame $Q$ is evolved by $J_t Q$ with a QR factorization at every step, accumulating $\tfrac1T\sum_t\log|\!\operatorname{diag}(R_t)|$ after a $300$-step burn-in over $3000$ steps, giving the sorted exponents $\lambda_1\ge\cdots\ge\lambda_N$. The same orbit and indexing as the top-exponent estimate of Experiment~1 are used, so $\lambda_1$ reproduces that value. We report $\lambda_1$, the summed spectrum, the operator norm, and the participation ratio of the hidden-state covariance, for a frozen reservoir (operator norm $0.8$) and a trained network on each driver.

Experiments~6 and~7 show that generalized synchronization and the embedding do not require training: a frozen reservoir already exhibits them. What, then, does training the recurrent weights change? We compare the conditional Lyapunov spectrum of a frozen reservoir and a trained network driven by the same signal. The conditional Lyapunov exponents are the per-step contraction rates of the fiber Jacobian $J_t=\operatorname{diag}(1-h_{t+1}^2)W$ along the driven orbit, computed by a QR iteration; the top exponent reproduces the power-iteration value of Experiment~1.

\begin{figure}[ht]
\centering
\includegraphics[width=0.9\textwidth]{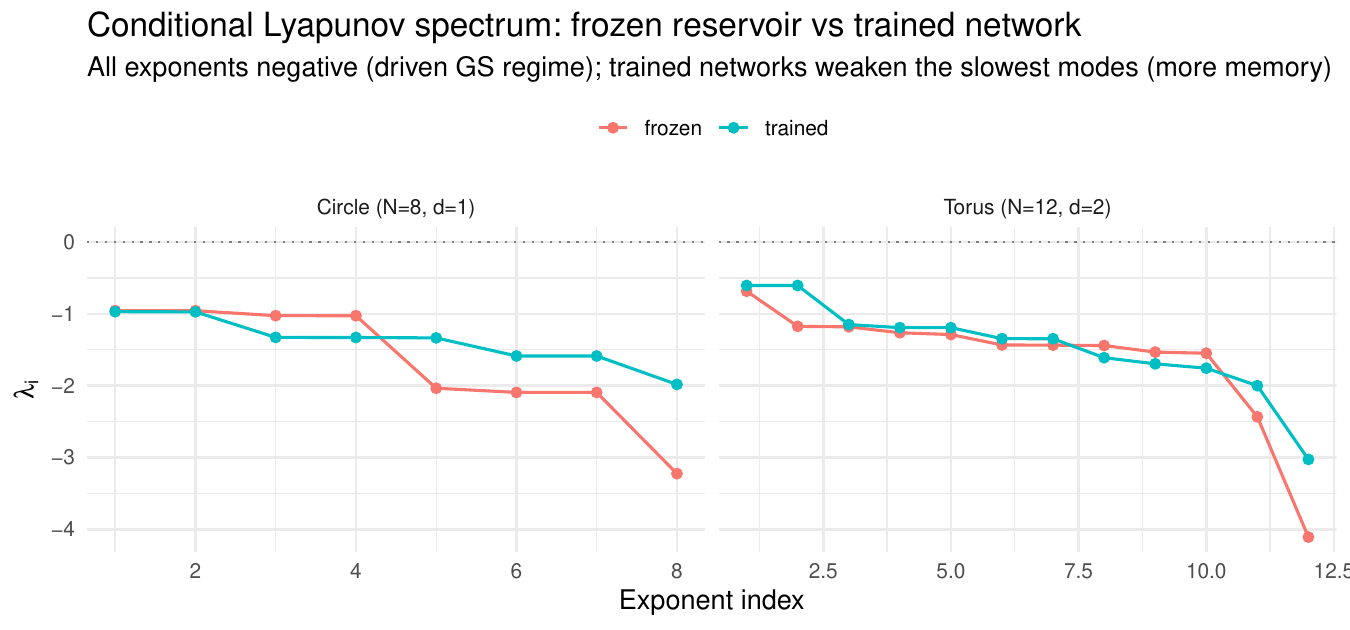}
\caption{Conditional Lyapunov spectrum of a frozen reservoir and a trained network, for the circle ($N=8$) and torus ($N=12$) drivers. Every exponent is negative (the driven synchronization regime). Training weakens the aggregate contraction (the spectrum lifts toward zero) and spreads contraction across more directions rather than speeding it; the slowest mode lifts toward zero for the torus but not the circle.}
\label{fig:spectrum}

\end{figure}

The spectrum stays strictly negative under training, so the network keeps synchronizing; training does not create generalized synchronization, and by raising the operator norm (from $0.80$ to $0.91$--$0.93$) it does not accelerate synchronization either. The slowest conditional mode, which sets the synchronization rate, is essentially unchanged for the circle ($-0.95$ to $-0.97$) and slower for the torus ($-0.68$ to $-0.60$), and re-synchronization time is comparable to the reservoir (Experiment~7). What training changes is the distribution of contraction: it weakens aggregate contraction (the summed exponents move toward zero, from $-13.4$ to $-11.1$ for the circle and from $-19.5$ to $-17.5$ for the torus) and broadens participation across hidden directions (the participation ratio rises from $1.14$ to $1.42$ for the circle and from $1.25$ to $1.58$ for the torus). Training moves a strongly contracting generic filter toward a less globally contractive, more distributed representation, and shapes that representation for the training objective rather than speeding synchronization. The robust, driver-independent statement is that training raises the operator norm, leaves synchronization no faster (slower for the torus), weakens aggregate contraction, and broadens spectral participation, while preserving a strictly negative spectrum (Figure~\ref{fig:spectrum}).
\par

\subsection{Experiment 9: Training Shapes Geometry for the Objective}
\paragraph{Methods.} Decodability compares three networks per driver (frozen reservoir, one-step-trained, and five-step-trained), five seeds each. The five-step trainer minimizes $\frac1{T}\sum_t\sum_{k=1}^{5}(\hat u_{t+k}-u_{t+k})^2$ with a five-output linear readout and backpropagation through time (the gradient is verified against finite differences before any run). Each network's hidden states drive a ridge readout (regularizer $10^{-3}$) fit on the first half of the run and evaluated on the second, for one-step and five-step targets; the nonlinear readout applies the same ridge to random tanh features (200 features, drawn once per driver and shared across the three networks). The embedding-intact check reports, on the same hidden states, the fifth-percentile normalized separation and the collision fraction. The class-specificity test (three seeds, $N\in\{5,8,12\}$) drives circle-trained, torus-trained, and reservoir networks with both signals and reports the fifth-percentile normalized hidden distance among distinct base-state pairs.

Training redistributes contraction (Experiment~8); does it also make the representation more useful? We measure decodability: how well a linear readout recovers future inputs from the hidden state, for a frozen reservoir, a network trained on one-step prediction, and a network trained on a five-step objective $\frac1T\sum_t\sum_{k=1}^{5}(\hat u_{t+k}-u_{t+k})^2$. All three receive the same ridge readout, fit on one half of the run and evaluated on the other; the only difference is how the recurrent weights were obtained.

\begin{figure}[ht]
\centering
\includegraphics[width=0.9\textwidth]{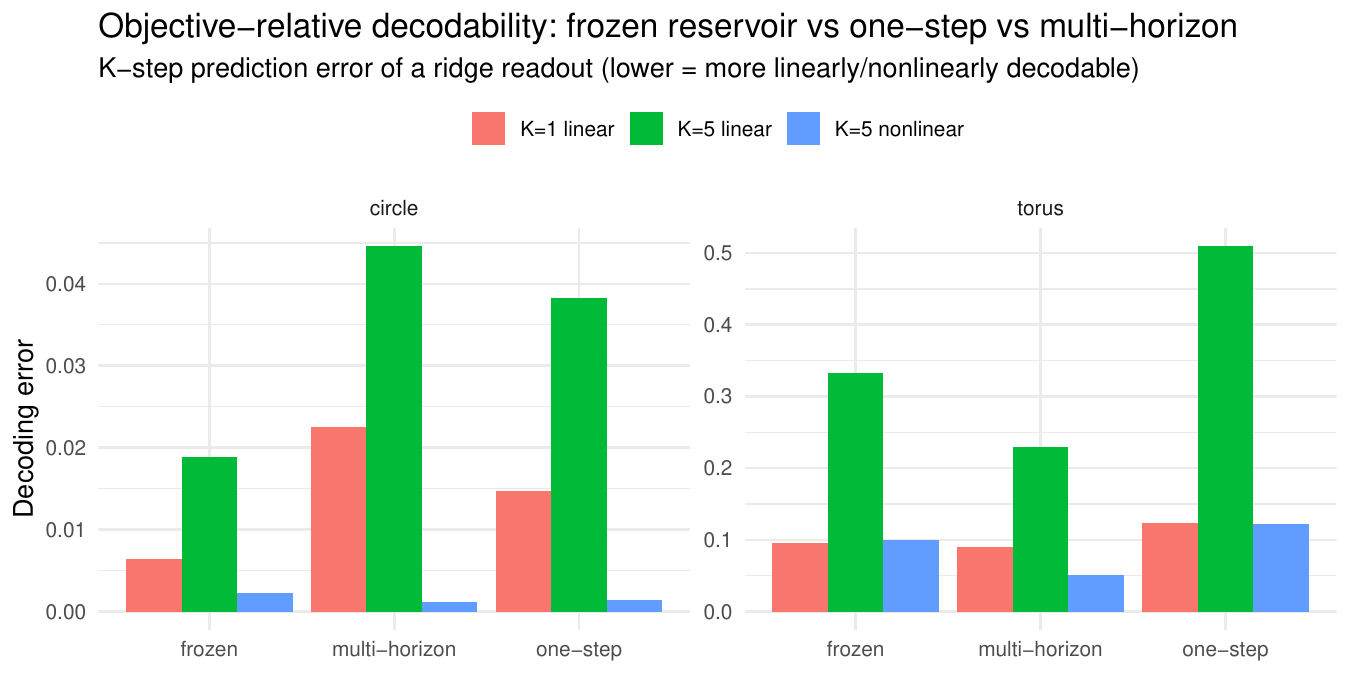}
\caption{Objective-relative decodability. K-step prediction error of a ridge readout (lower is better) for a frozen reservoir, a one-step-trained network, and a five-step-trained network, for the circle ($N=8$) and torus ($N=12$) drivers. On the torus, one-step training makes five-step linear decoding worse than the frozen reservoir; the five-step-trained network recovers it. The nonlinear (random-feature) readout closes the gap, showing the information is present but less linearly accessible.}
\label{fig:decode}

\end{figure}

A frozen reservoir is often the better linear feature space. For the circle it gives the lowest one-step and five-step linear error (Figure~\ref{fig:decode}); training the recurrent weights does not improve linear decodability there. For the torus, one-step training makes five-step linear decoding worse than the reservoir ($0.51$ versus $0.33$): optimizing one-step prediction warps the geometry away from what a linear five-step readout needs. Training on the five-step objective recovers it ($0.23$, below the reservoir's $0.33$). The geometry follows the loss horizon.

The warped information is not destroyed: a nonlinear readout (random features) recovers the five-step accuracy ($0.05$ for the five-step-trained torus network), and the embedding itself is intact, with worst-case state separation no worse than the reservoir's and collision fractions no higher. Training thus changes which futures are linearly accessible without breaking the embedding, so prediction error, embedding quality, and linear accessibility are distinct quantities.

\begin{figure}[ht]
\centering
\includegraphics[width=0.9\textwidth]{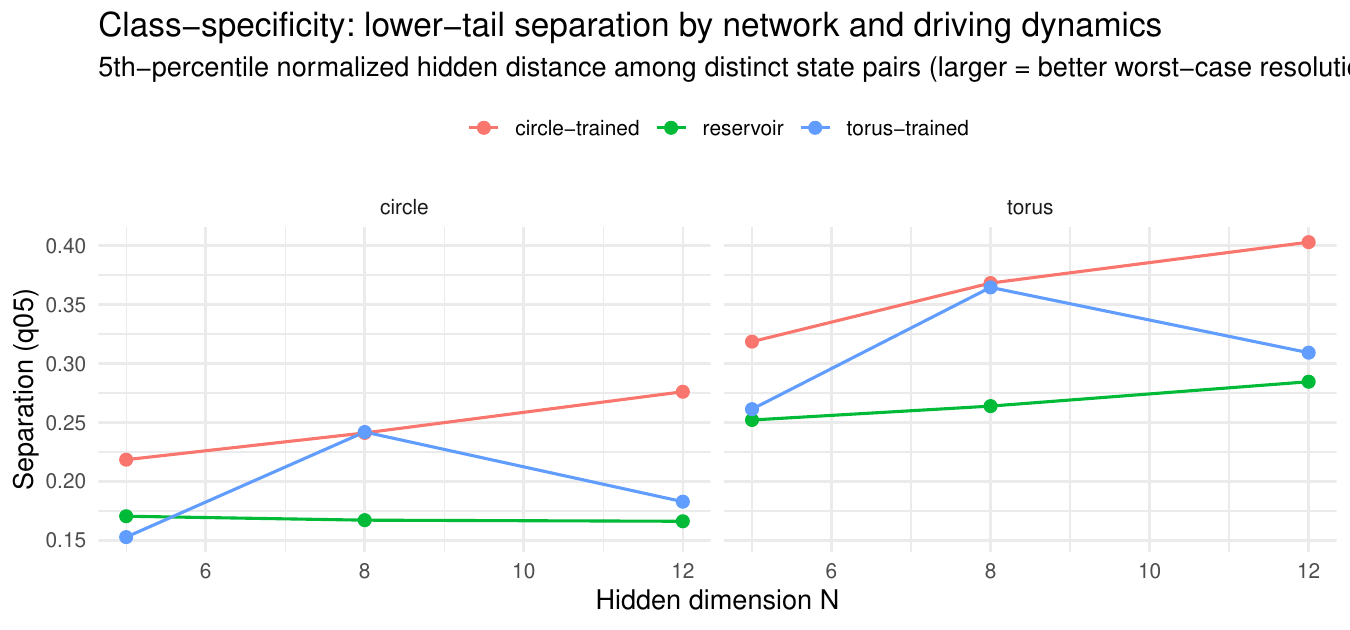}
\caption{Class-specificity test (inconclusive). Worst-case separation (fifth-percentile normalized hidden distance among distinct states, larger is better) for circle-trained, torus-trained, and reservoir networks against each driver, by hidden dimension. Training improves separation over the reservoir, but the circle-trained network separates best on both drivers and the torus-trained network does not reliably exceed the reservoir on the circle: the geometry is not strongly specialized to the trained dynamics.}
\label{fig:classspec}

\end{figure}

A companion test for task-specific geometry was inconclusive. Driving circle-trained, torus-trained, and reservoir networks with both signals and measuring worst-case separation (the fifth-percentile normalized hidden distance among distinct states), training improved separation over the reservoir but not in a class-specific way: the circle-trained network separated best on both drivers, and the torus-trained network did not reliably exceed the reservoir on the circle (Figure~\ref{fig:classspec}). At these hidden dimensions the embedding is not strongly specialized to the dynamics it was trained on.
\par

\end{appendices}

\bibliography{refs}

\end{document}